\documentclass[11pt,a4paper]{article}
\usepackage{my-own-commands}

\title{CMB in the river frame and gauge invariance at second order}

\author{Omar Roldan}

\affiliation{Instituto de F\'\i sica, Universidade Federal do Rio de Janeiro, 21941-972, \\ 
Rio de Janeiro, RJ, Brazil}

\emailAdd{oaroldan@if.ufrj.br}

\abstract{
GAUGE INVARIANCE: The Sachs-Wolfe formula describing the Cosmic Microwave Background (CMB) temperature anisotropies is one of the most important relations in cosmology. Despite its importance, the gauge invariance of this formula has only been discussed at first order. Here we discuss the subtle issue of second-order gauge transformations on the CMB. By introducing two rules (needed to handle the subtle issues), we prove the gauge invariance of the second-order Sachs-Wolfe formula and provide several compact expressions which can be useful for the study of gauge transformations on cosmology. Our results go beyond a simple technicality: we discuss from a physical point of view several aspects that improve our understanding of the CMB. We also elucidate how crucial it is to understand gauge transformations on the CMB in order to avoid errors and/or misconceptions as occurred in the past. THE RIVER FRAME: we introduce a cosmological frame which we call the river frame. In this frame, photons and any object can be thought as fishes swimming in the river and relations are easily expressed in either the metric or the covariant formalism then ensuring a transparent geometric meaning. Finally, our results show that the river frame is useful to make perturbative and non-perturbative analysis. In particular, it was already used to obtain the fully nonlinear generalization of the Sachs-Wolfe formula and is used here to describe second-order perturbations.
}


\begin{document}
\maketitle

\section{Introduction and main results}

\subsection*{Gauge invariance}

Gauge dependence and gauge invariance are crucial concepts in cosmological perturbation theory. While many quantities often used in cosmology are gauge dependent (e.g. the metric perturbations, density perturbations, velocity perturbations, etc.), physical observables must be expressed in terms of gauge-invariant quantities. One of the most important observables in cosmology is provided by the Cosmic Microwave Background (CMB) temperature anisotropies.
In the instant decoupling approximation%
\footnote{In this approximation, all the CMB photons which are observed today where emitted at the same time.
} %
and neglecting secondary scatterings,%
\footnote{That is, scatterings of CMB photons with hot gas during its way down to the observer.
} %
such anisotropies are described by the so called Sachs-Wolfe formula \cite{Sachs:1967er,Roldan:2017wvm}. Of particular importance is the \textit{second-order Sachs-Wolfe formula} which describes in a unified picture the Sachs-Wolfe and the integrated Sachs-Wolfe (and Rees-Sciama) effects \cite{Rees:1968zza,Crittenden:1995ak}, lensing \cite{Seljak:1995ve,Zaldarriaga:1998ar}, time-delay \cite{Hu:2001yq}, Doppler modulation and aberration \cite{Roldan:2017wvm,Roldan:2016ayx}. It is also necessary for a full characterization of the three-point function (or its Fourier counterpart, the bispectrum) which is a very important tool for the search of primordial non-Gaussianity \cite{NG:Ade:2015ava}. Other useful references are \cite{Boubekeur:2008kn,Bartolo:2005fp,Bartolo:2005kv,Bartolo:2006cu,Bartolo:2006fj,Pitrou:2008hy,Senatore:2008vi}.

Although it is clear that the Sachs-Wolfe formula has to be gauge invariant (because it provides an observable quantity), studying its gauge independence and in general, studying how gauge transformations should be applied to the CMB is important because it improves our understanding of the CMB and our ability to make predictions. It also minimizes the chances for making mistakes in theoretical descriptions. To illustrate this, we give an example. In a nice article, Creminelli et al.~\cite{Creminelli:2011sq} obtained the squeezed limit of the CMB bispectrum. The result and the method are very simple an elegant: by arguing that a superhorizon perturbation (coming from adiabatic initial conditions during single-field-inflation) is locally equivalent to a coordinate (or gauge) transformation, they obtained a formula that is supposed to include corrections when the long-mode reenters the horizon. However, their basic formula (Eq.27) has an error as was noted (and corrected) in \cite{Mirbabayi:2014hda}. The error appears because making gauge transformations on the CMB is non trivial (see \refs{sec:GT-X} for more discussion on this).

At first order, the gauge independence of the temperature anisotropies $ \Theta$ was discussed in some works, either by expressing $ \Theta$ in terms of fully gauge-invariant quantities \cite{stoeger1995gauge,Dunsby:1997xy,Challinor:1998aa,Hwang:1999fj}, or by showing this explicitly \cite{Zibin:2008fe}. All these works have in common the so called \textit{covariant formalism}. Alternatively, a very straightforward proof was given in (section 3.3 of) \cite{Roldan:2017wvm} by using the metric formalism and the concept of intrinsic temperature anisotropies $ \cT$. Regarding the second-order $ \Theta$, an explicit proof of its gauge invariance does not appear to have been presented before (apart from the analysis of \cite{Bartolo:2004ty}). In this work we address such a problem and point out some subtle issues behind the proof, giving special emphasis on the physics behind these subtle issues.

\subsection*{The river frame}


To discuss the gauge transformations on the CMB, we will use the results of \cite{Roldan:2017wvm} where the fully nonlinear Sachs-Wolfe formula was obtained. These results were obtained thanks to the introduction of a cosmological frame (an orthonormal basis) which highly simplified the calculations. Here we will refer to this frame as \textit{the river frame}. The name is motivated by an analogy with \textit{The river model of black holes} \cite{Hamilton:2004au} in which space itself flows like a river through a flat background, while objects (fishes) move through the river according to the rules of special relativity. The river model of black holes emerged thanks to the works of Gullstrand~\cite{Gullstrand:1922tfa} and Painlev\'e~\cite{Painleve:1921} who realized that the Schwarzschild metric can be expressed in the form
\begin{align*}
	\dd s^{2} = - \dd t_{ff}^{2} + \lp \dd r + \beta_{r}\, \dd t_{ff} \rp^{2} + 
	r^{2} \lp \dd \theta^2 + \sin^{2}{\theta}\ \dd \phi^{2}\rp  \,,
\end{align*}
where $ \beta_{r} = \sqrt{2GM/r}$ is the Newtonian escape velocity at a radius $ r$ and $ \dd t_{ff}$ is the (differential) proper time experienced by an object that free falls radially inward from zero velocity at infinity. As nicely described in \cite{Hamilton:2004au} and \cite{Hamilton:2017}, the Gullstrand-Painlev\'e metric give us a picture of space itself flowing like a river (with velocity $ \beta_{r}$)%
\footnote{In a proper time $ \dd t_{ff}$, the river moves a proper distance $ \dd r =  - \beta_{r}\ \dd t_{ff}$ through the background.
} %
 into the Schwarzschild black hole, and photons (and any other object) as fishes swimming in the river. This picture of a river of space is very useful to get a better and intuitive understanding of the physics of the Schwarzschild spacetime. 

After a coordinate transformation, the Gullstrand-Painlev\'e metric can be written in Cartesian form as
\begin{align}
		\dd s^2 = - \dd t_{ff}^{2} + \delta_{ij} \lp \dd x^{i} - \beta^{i}_{r}\ \dd t_{ff}\rp 
	\lp \dd x^{j} - \beta^{j}_{r}\ \dd t_{ff}\rp  \,,
	\label{eq:gull-metric-intro}
\end{align}
where $ \beta^{i}_{r} = \beta_{r}\, x^{i}/r$ and $ \delta_{ij}$ is the Kronecker delta symbol. \eq{eq:gull-metric-intro} is our starting point for what we call the cosmological river frame.



\subsection{Main results}

All our results are based in the metric parametrization (introduced in \cite{Roldan:2017wvm})
\begin{align}
	\dd s^2  &  = a^2(\eta) e^{2\Phi} \lc - \dd \eta^2 + 
	2 \beta_{j} \lp e^{-M} \rp^{j}_{\ i}\ \dd x^i \dd \eta 
	+ \lp  e^{-2M}\rp_{ij}\dd x^i \dd x^j \rc \,.
	\label{eq:metric}
\end{align}
Here $ M$ is a symmetric matrix and the indices of $ \beta^{i}$ and $ M_{ij}$ are lowered and raised with a Kronecker delta. We can think of $ \Phi, \beta^{i}$ and $ M$ as the nonlinear versions of the usual metric perturbations around the FLRW Universe. As discussed in \refs{sec:river}, this metric can be written in a way which mimics the Gullstrand-Painlev\'e Cartesian metric \eq{eq:gull-metric}. There are however significant differences as we point out in \refs{sec:river-black holes}.

Regarding the gauge transformations, in this work we will use the \textit{active approach}. That is, we transform the fields (by using Lie derivatives) while keeping the spacetime coordinates unchanged. This will be important when applying gauge transformations to the CMB. For instance, the definition of the the last scattering surface remains simple even after applying a gauge transformation. Indeed, if choose a coordinate system such that the last scattering surface is placed at a fixed (conformal) time of emission $ \eta_e$,%
\footnote{In this coordinate system all CMB photons are emitted at the same conformal time $ \eta_e$ independently of the direction of observation. Such coordinate system should exist if we assume the instant decoupling approximation as we do here. 
} %
then after a gauge transformation the coordinates of the spacetime remains the same and so the last scattering surface still belongs to the hypersurface of $ \eta = \eta_e = const$. This is not the case however if we use the passive approach to gauge transformation (or coordinate transformations) for which $ \eta \to \hat{ \eta} (\eta, x^{i})$ (see \cite{Mirbabayi:2014hda} for an specific example).

In \refs{sec:gt2nd} (see also Appendix~\ref{app:2ndGT}) we provide a set of second-order gauge transformations of cosmological interest putting special emphasis on the gauge transformations on the CMB (in \refs{sec:GT-nixi}). To get a feeling of our results, we plug here a few.

Our notation is as follows: after a gauge transformation a geometrical object T (e.g., the metric, a vector field, the coordinates of the photon's trajectory, etc.) transform as $ T \to T_{*} $. In particular, up to second order we have 
\begin{align}
	\Phi_{*} & = \lc \Phi + \alpha' +\cH \alpha \rc  + 
	\f12 \xi^{c} \p{c} \lp \Phi + \Phi_{*} \rp - \f12 \lp \beta + \beta_{*} \rp^{i} (\xi^{i})'\,, 
	\label{eq:GT-phi-intro}\\
	\cT_{*} & = \cT - \cH \alpha + \f12 \xi^{c} \p{c} \lp \cT + \cT_{*} \rp \,, 
	\label{eq:GT-cT-intro}
\end{align}
where $ \cT$ is the (logarithmic) intrinsic temperature perturbations (see next section). Similar expressions are provided for $ \beta_{*}^{i}$ and all other geometrical objects. Here $ \xi^{\mu} = (\alpha,\xi^{i})$ is the vector field generating the gauge transformation, and a ``prime'' means derivative w.r.t conformal time.

Note that the first-order relations are simply obtained by neglecting the explicitly quadratic terms. For instance, at first order we have: $ \Phi_{*} = \lc \Phi + \alpha' +\cH \alpha \rc $. By using the first-order relations we can easily rewrite the second-order ones, e.g., 
$ \cT_{*} =  \cT - \cH \alpha + \xi^{c} \p{c} \lp \cT- \cH \alpha/2 \rp$. So our expressions are useful for recursive computations. Additionally, by writing the gauge transformations in the form given above, we can see which variables are directly related by the gauge transformations. This can be useful for instance in constructing gauge-invariant objects. 

Although in the active approach the coordinates of the spacetime are fixed, this is not the case for the coordinates $ x^{i}( \eta)$ of the photon's geodesic. The reason is simple: gauge transformations act on the metric field, and the photon's path depends on the metric. This fact turns out to be very relevant and makes the gauge transformations of the CMB non trivial. In \refs{sec:GT-nixi} we show that the coordinates of the photon's path transform up to first order as
\begin{align}
	\eta_{*} = \eta \,, \qquad x^{i}_{*} = x^{i} - \lp \alpha\ \un{\tn}{i} + \xi^{i} \rp  \,,
	\label{eq:x*}
\end{align}
where $ \un{\tn}{i}$ is the direction vector, which at zero order defines the radial direction. Compare this result with the standard transformation of coordinates $ x^{a}_{*} = x^{a} - \xi^{a}$. Because of the previous relations, we argue (in \refs{sec:GT-nixi}) that when applying gauge transformations to the CMB a ``rule'' should be used%
\footnote{This is \texttt{Rule 1}, a related \texttt{Rule 2} is given in \refs{sec:GTisw}.
} %
\begin{align}
	\texttt{Rule 1:} \qquad \qquad 
	\xi^{c} \p{c} \to \lp \alpha\ \un{\tn}{i} + \xi^{i} \rp \p{i} \,,
	\label{eq:rule1}
\end{align}
that is, the time derivative $ \p{0}$ is changed by a radial derivative, $ \p{r} = \un{\tn}{i} \p{i}$. Physical arguments which justify the use of this rule are provided in \refs{sec:GT-X}. A rigorous proof of such a rule is left for a future work.

Finally, by taking into account these transformations and using the prescribed rules, we prove the gauge invariance of the second-order Sachs-Wolfe formula in \refs{sec:GInvariance}. Throughout the article we will use $ a, b, c, \cdots$ and also Greek letters to represent the spacetime indices and $ i, j, k, \cdots$ to represent the spatial indices. Quantities defined w.r.t the river frame will have a ``tilde'', for example $ \tn, \tb, \cdots$.

\section{Generalized Sachs-Wolfe formula: quick review}
\label{sec:GSW}


In a recent paper \cite{Roldan:2017wvm}, an exact expression for the observed CMB temperature was obtained.%
\footnote{The result assumes that the CMB spectrum is blackbody, that is, no spectral distortions are included.
} %
The result is valid at all orders in perturbation theory, is also valid in all gauges and includes scalar, vector and tensor modes. In this section we quickly review the main results of \cite{Roldan:2017wvm} and in \refs{sec:CMB2nd} we specialize to the second-order case. The results of \cite{Roldan:2017wvm} state that the observed CMB temperature can be written as $ T_o = \bar{T}_o\ e^{\Theta}$, where $ \bar{T}_o$ is the observed mean temperature and 
\begin{align}
	%
	\Theta & \equiv (\cT_e - \cT_o) + (\Phi_e - \Phi_o)  + I_0 + 
	\ln \lp \f{\gamma_e \lp 1 - n_e \cdot v_e \rp}{\gamma_o \lp 1 - n_o \cdot v_o \rp} \rp 
	  \,,
	\label{eq:Gsw-intro} \\
	I_{0} & \equiv  \int_{ \eta_e}^{ \eta_o} \dd \eta\ \lc \f{ \tb \cdot \tb'}{1- \tb^{2}} + 
	\f{ \tn \cdot \tb' + \tn \cdot A_{0} \cdot \lp \tn + \tb \rp }{1 + \tb \cdot \tn}\rc \,.
	\label{eq:I0}
\end{align}
The subscripts $ e$ and $ o$ mean that quantities must be evaluated at the emission and observation event respectively.
Here $ \cT$ and $ I_0$ are respectively, the nonlinear generalization of the intrinsic temperature anisotropies and the integrated Sachs-Wolfe effect (ISW). The logarithm term corresponds to the Doppler effect, with $ \gamma$ the Lorentz factor, $ v_o$ (and $ v_e$) are the peculiar velocity of the observer (and emitter), $ n_o$ the direction of observation and $ - n_e$ the direction of emission. 
\OLD{$ x_e = ( \eta, x^{i})_e$ and $x_o = ( \eta, x^{i})_o $ are the spacetime coordinates of the emission and observation events.
}
Regarding the ISW term, the integration must be performed along the photon's (curved) path $ x^{\mu} ( \eta)$. The emission time $ \eta_e$ need not be the same for all the directions of observations. Although in most cases it is useful to choose the coordinate system in such a way that the last scattering surface is located on the hypersurface of $ \eta = \eta_e  = const$.

The precise physical meaning of $ \tn$ and $ \tb$ is given in \refs{sec:river}. For the impatient reader we can say that: $ - \tn$ and $ \tb$ are respectively the direction of propagation of CMB photons and the (relative) velocity of comoving observers as seen in the river frame. Explicitly, we have%
\footnote{The definition of $ A_{0}$ follows from the Baker-Campbell-Hausdorff formula (or the Zassenhaus formula) \cite{Wilcox:1967zz}.
} %
\begin{align}
	\un{ \tb}{a}  & \equiv \lp 0, \beta^{i}/\beta^{0} \rp \,,  
	&  & \textrm{with}  &  & \beta^0 \equiv \sqrt{1 + \beta_i \beta^i}\,, \\
	A_{0} & \equiv \int_{0}^{1} \dd s\ e^{- s M} \lp \p{0} M \rp e^{s M} \,,
	&  & \Longleftrightarrow  
	&  &  \lp \p{0}\, e^{-M} \rp \equiv - A_{0}\, e^{-M} \,.
	\label{eq:Adef}
\end{align}
In Eqs.~\eqref{eq:Gsw-intro}-\eqref{eq:I0}, a ``$ \cdot$'' represents scalar product between tensors, that is, $ v \cdot n  = v^{a} n_{a} = \un{v}{a}\ \unI{n}{a}$. A bar on the indices is used to distinguish coordinates components (in the background frame) from the tetrad components (in the river frame). In this sense the symmetric matrix $ (A_{0})_{ij}$ can be thought as the non-vanishing components of a (space-like) tensor $ A_{0}$ in the river frame, that is: $ \unI{(A_{0})}{ij} \equiv (A_{0})_{ij}$ and $ \unI{(A_{0})}{0 a} = 0$.

A direct consequence of \eq{eq:Gsw-intro} is that the maps of the \textit{logarithmic temperature anisotropies} $ \Theta = \ln \lp 1 + \Delta T_o/ \bar{T}_o \rp $ are much cleaner than the usual CMB maps of $ \Delta T_o/ \bar{T}_o$. This follows for instance, from the fact that the dependence of $ \Theta$ on $ \cT$ and $ I_0 $ is linear, while the temperature anisotropies
\begin{align}
	\f{ \Delta T_o}{ \bar{T}_o} \equiv e^{\Theta} - 1 = \Theta + \f{\Theta^{2}}{2} + \cdots\,,
	\label{eq:DT2}
\end{align}
contain terms involving products of these quantities, then correlating the ISW with the intrinsic anisotropies. So using maps of $ \Theta$ can for instance, help to disentangle the nonlinear ISW from other effects and facilitate the search for primordial non-Gaussianity.

Note that in \eq{eq:Gsw-intro} the Doppler effect is given in terms of the peculiar velocity $ v$ (of the observer and emitter) and the direction of observation $ n$ in the frame of the observer and emitter. We can express \eq{eq:Gsw-intro} fully in terms of the quantities defined w.r.t the river frame by using
\begin{align}
	\gamma \lp 1 - n \cdot v \rp  = 
	\f{ \tilde{\gamma} \lp 1 + \tn \cdot \tb \rp}{\tilde{\gamma} \sub{F} \lp 1 + \tn \cdot \vf \rp}	 \,.
	\label{eq:Doppler-river}
\end{align}
Here, if $ v$ is the peculiar velocity of the observer (emitter), then $ \vf$ is the velocity of the observer (emitter) w.r.t the river frame, and $ \tilde{\gamma} \sub{F} = \lp 1 - \vf \cdot \vf \rp^{- 1/2}$. The subscript $ F$ is because we think of objects as fishes moving through the river.

\subsection{The (logarithmic) intrinsic temperature anisotropies}
\label{sec:cT}

We now review the definitions of the \textit{(logarithmic) intrinsic temperature anisotropies} $ \cT_e$ and $ \cT_o$. This latter factor which is absent in previous works in literature is important for two reasons: it makes the expression for $ \Theta$ symmetric in the emission and observation points and it ensures the gauge invariance of $ \Theta$.

Until decoupling time (for $ \eta \le \eta_e$) photons are locally in thermal equilibrium with baryons, forming the so called photon-baryon fluid. We write the temperature of this fluid as
\begin{align}
	T( x, n) = \braket{ T}\, e^{ \cT} \,, 	\qquad	\textrm{where} \qquad \cT = \cT( x, n)\,,
\end{align}
%
and $ \braket{T} \propto 1/a(\eta)$ is the background temperature ($ a = a( \eta)$ is the scale factor). This expression is only defined for $ \eta \le \eta_e$. The extension of $ \cT$ for $ \eta > \eta_e$ is given below. For the emission points we have $ T_{e} = \braket{T}_{e} e^{ \cT_e }$. We stress that the mean $ \braket{}$ is taken on the space-like 3D-hypersurfaces of constant $ \eta$, however, what is important for observations%
\footnote{The relevance of this fact was also pointed out in some previous works, see for instance \cite{Mirbabayi:2014hda} and after Eq.(10) of \cite{Hwang:1999fj}. Note also that this section has some overlap with what is discussed after Eq.(3) of \cite{Bartolo:2004ty}.
} %
is the mean taken on the last scattering surface $S_{e,o}$ (mean values on $S_{e,o}$ represent integrations w.r.t the direction of observation $ n_{o}$).

Although our results are independent of the specific way in which we define $ S_{e,o}$, we remind the reader that in the instant decoupling approximation we can find a coordinate system in such a way that all CMB photons were emitted at a fixed conformal time $ \eta_e$.%
\footnote{Note that we can choose the hypersurfaces of $ \eta = const$ to coincide with those of $ \braket{T} = const$. This is always possible as far as the spacetime can be foliated by space-like hypersurfaces (globally hyperbolic spacetime). See the $ 3 + 1$-formalism, section 3 of \cite{Gourgoulhon:2007ue}.
} %
In that sense, we can think of $S_{e,o}$ as the 2D-surface formed at the intersection between the hypersurface of $ \eta = \eta_e$ and the past light cone of the observer.

\OLD{In the instant decoupling approximation we can think of the CMB photons as being emitted at a fixed time $ \eta_e$, any emission point having a temperature $ T_{e} = \braket{T}_{e} e^{ \cT_e }$. We stress that the mean $ \braket{}$ is taken on the space-like 3D-hypersurfaces of constant $ \eta$. 
However, what is important for observations is the mean taken on the last scattering surface%
\footnote{The relevance of this fact was also pointed out in some previous works, see for instance \cite{Mirbabayi:2014hda} and after Eq.(10) \cite{Hwang:1999fj}.
} %
$S_{e,o}$, defined  as the 2D-surface formed at the intersection between the hypersurface of $ \eta = \eta_e$ and the past light cone of the observer. It follows from $ T_o = \bar{T}_o\ e^{\Theta}$, that $ \overline{ \exp \lp \Theta \rp} = 1$, where we use an overline to denote mean values on $S_{e,o}$. Then from \eq{eq:Gsw-intro} we have
}

It follows from $ T_o = \bar{T}_o\ e^{\Theta}$, that $ \overline{ \exp \lp \Theta \rp} = 1$, where we use an overline to denote mean values on $S_{e,o}$. Then from \eq{eq:Gsw-intro} we have
\begin{align}
	e^{\cT_o} = \overline{ \exp \lp { \cT_{e} + \Phi_e - \Phi_o + I_{0} + 
	\ln \gamma_o \lp 1 - n_o \cdot v_o \rp - \ln \gamma_e \lp 1 - n_e \cdot v_e \rp} \rp} \,.
	\label{eq:cT_o}
\end{align}
In fact, this was the definition of $ \cT_o$ that we gave in \cite{Roldan:2017wvm}, and with that we obtained \eq{eq:Gsw-intro}. It can be shown \cite{Roldan:2017wvm} that $ \cT_o$ is related to the observed mean temperature by
\begin{align}
	\bar{T}_o( x_{o}) & = \f{a_e}{a_o} \braket{T}_{e} e^{ \cT_o} \,.
	\label{eq:barTo}
\end{align}
Since $ \braket{T}_{e} \propto 1/a_{e}$, it follows that $ \cT_o = \cT_o ( x_{o})$ transforms under gauge transformations in the same way as the logarithmic anisotropies $ \cT_{e}$ but evaluated at the observer's position. Since this definition is valid for any observer with $ \eta_o  > \eta_e$, it provides a natural extension for the field $ \cT$ to the whole spacetime. Note however that by construction $ \cT_o$ depends only on the spacetime position $ x_{o}$ not on the direction of observation $ n_{o}$. This is in contrast with intrinsic temperature anisotropies $ \cT_e$ which in general depends on both $ x_{e}$ and $ n_{e}$. Indeed, the intrinsic temperature anisotropies have a quadrupole component which act as a source for the CMB polarization \cite{Rees:1968,Ferte:2015oga}. From \eq{eq:barTo} it seems as if $ \cT_o$ should also depend on $ \eta_e$. However, this cannot be the case, because for a given observer (at the spacetime position $ x_{o}$) there is one and only one last scattering surface. So we cannot think of $ \eta_e$ as a free parameter for $ \cT_o$.

Note also that $ {\mathcal T}_o$ is a monopole term which varies according to the observer, and which contributes to the definition of the mean temperature according to \eq{eq:barTo}.%
\footnote{I thank the anonymous referee for drawing it to my attention.} %
It is this fact that implies that a constant scalar mode have no observable effect on the (second-order) temperature anisotropies, as was pointed out in \cite{Boubekeur:2009uk} (see after their Eq.(2.33)) so confirming the results of \cite{Bartolo:2005fp,Bartolo:2004ty,Bartolo:2003gh,Bartolo:2003bz}. For comparison with our work, note that the terms $ \Phi_{e}/3$ and $ \braket{ e^{ \Phi_{e}/3}}$ in Eq.(2.34) of \cite{Boubekeur:2009uk} are just our $ \Phi + \cT$ and $ e^{\Phi_{o} + \cT_o}$ in the limit of large scales and for adiabatic perturbations.

\begin{figure}[h!]\centering
	\includegraphics[scale=0.35]{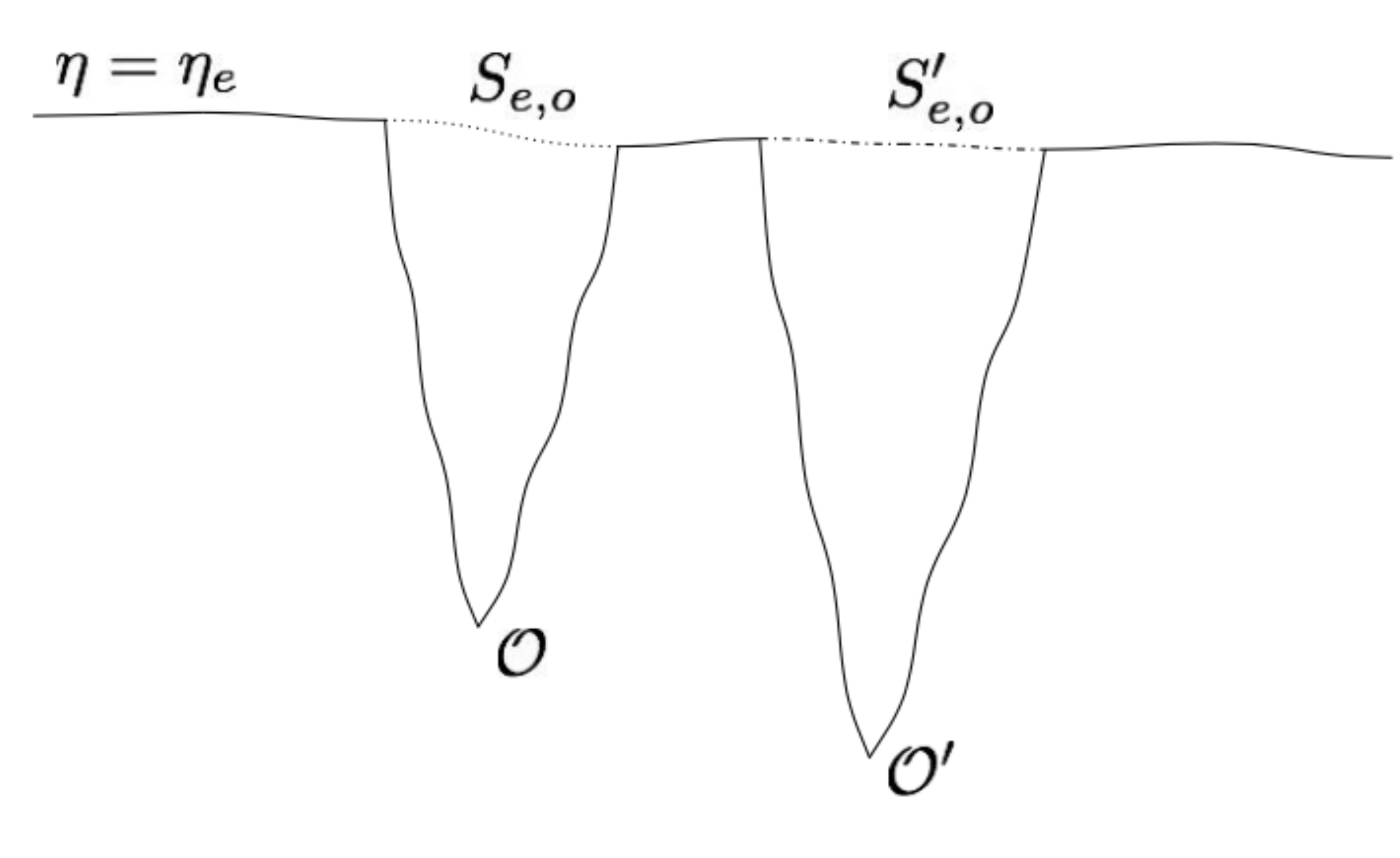}
	\caption{Different observers define different hypersurfaces $S_{e,o}$ each one with its own mean value temperature $ \bar{T}_{e}$. The deviation of $ \bar{T}_{e}$ from the mean temperature at the hypersurface $ \eta = \eta_e$ will therefore depend on the observer's position $ x_{o}$, and that information is encoded into $ \cT_o$.}
	\label{fig:lss}
\end{figure}

Let's define the mean temperature $ \bar{T}_e$ of the last scattering surface as (this definition is meaningful only if we choose the coordinate system such that $S_{e,o}$ is located in the $ \eta = \eta_e = const$ surface)
\begin{align}
	\bar{T}_e( x_{o};\eta_e) \equiv \braket{T}_{e} e^{ \cT_o} \,.
\end{align}
Because in general $ \bar{T}_e \ne \braket{T}_{e}$, then through the previous equation, $ \cT_o$ tell us how anisotropic the last scattering surface is (see figure~\ref{fig:lss}). We stress that unlike $ \braket{T}_e$, it is $ \bar{T}_e$ that is the relevant quantity to describe the observed temperature. In particular the observed monopole of the CMB, $ \bar{T}_{o}$, is related to the monopole at the last scattering surface $ \bar{T}_{e}$ by the relation $ \bar{T}_o = \bar{T}_e\ a_e/a_o$.

For completeness (although not necessary for the purpose of the present paper), let's discuss how the intrinsic anisotropies $ \cT$ are related to the photon energy density anisotropies. We will write any scalar field $ S$ as: $ S = \braket{S} e^{ \sigma}$, where $ \braket{S}( \eta)$ is the background value and $ \sigma$ gives the logarithmic anisotropies of $ S$. In particular, for the energy density we write
\begin{align}
	\rho \equiv \braket{ \rho} e^{ \delta}\,.
	\label{eq:rho_delta}
\end{align}
This notation is very convenient as some relations that are valid at first-order will also hold at all orders. For instance, it is well known that at first order we have $ \cT = \delta_{ \gamma}/4$, where $ \delta_{ \gamma}$ is the density contrast of photons (see for instance \cite{Bartolo:2003bz} after their Eq.(79)). Now, we will show that in our notation, this is an exact result (valid at all order in perturbation theory). Indeed, since the energy density $ \rho_{ \gamma}$ of photons is related to its temperature $ T$ by $ \rho_{ \gamma} \propto T^{4}$ we obtain $ e^{ \delta_{ \gamma}} = e^{ 4 \cT}$\,, or 
\begin{align}
	\cT = \frac{\delta_\gamma}{4} = \f14 \ln \lp \frac{\rho_\gamma}{\braket{ \rho_\gamma}} \rp \,.
\end{align}
One additional example is provided by the isocurvature (or entropy) perturbation between two fluids $ A$ and $ B$, which is defined as $ S_{A,B} = 3 ( \zeta_{ A} - \zeta_{ B})$, with $ \zeta_{A,B}$ the curvature perturbation of the fluids $ A$ and $ B$ (see for instance \cite{Langlois:2011zz,Malik:2004tf}).%
\footnote{Take care that the definition of $ \zeta$ is different in both articles. See after Eq.(26) of \cite{Langlois:2011zz} for more details.
} %
It is well known that at first order we have
\begin{align}
	S_{A,B} = \frac{1}{1 + w_{A}} \delta_{A} - \frac{1}{1 + w_{B}} \delta_{B}\,,
	\label{eq:entropy}
\end{align}
where $ w_{A} \equiv P_{A}/\rho_{A} = const$, and $ P_{A}$ is the pressure of the fluid. For radiation $ w_{A} = 1/3$ and for a pressureless fluid $ w_{A} = 0$. By using the notation introduced in \eq{eq:rho_delta}, it follows that \eq{eq:entropy} is also an exact result (valid at all order in perturbation theory), as can be obtained directly form Eq.(8) of \cite{Langlois:2011zz}, see also \cite{Langlois:2008vk} (after Eq.(8))
and \cite{Bartolo:2005fp} (after Eq.(10)).

\section{The river frame}
\label{sec:river}

In this section we introduce a cosmological frame which we called the river frame. The idea of a river is borrowed from the work of Hamilton \& Lisle~\cite{Hamilton:2004au} that describes observers falling into a stationary black hole as being fishes falling in a waterfall (the waterfall being the river of space). Below we will use some basic notions of tetrads, for the unfamiliar reader we refer to the nice book \cite{Hamilton:2017} which describes in great detail, not only the concept of tetrads but also the river model for black holes. See also \cite{Ellis:1998ct,Carroll:1997ar} for more on tetrads.

The generalized Sachs-Wolfe formula \eq{eq:Gsw-intro} was obtained by writing the line element as $ \dd s^2  = a^2(\eta) e^{2\Phi} \dd \hat{s}^2$, with the conformal metric
\begin{align}
	\dd \hat{s}^2  & = - \dd \eta^2 + 
	2\beta_{j} \lp e^{-M} \rp^{j}_{\ i} \ \dd x^i \dd \eta 
	+ \lp  e^{-2M}\rp_{ij}\dd x^i \dd x^j \notag \\
	 & = - \lp \beta^0\dd \eta \rp ^2 + 
	\lc \lp e^{ - M} \rp ^{j}_{\ i} \dd x^i + \beta^j \dd \eta\rc 
	\lc \lp e^{ - M} \rp_{ jk} \dd x^k + \beta_j \dd \eta\rc \,,
	\label{eq:Cmetric}
\end{align}
and we have introduced $ \beta^0 \equiv \sqrt{1 + \beta_i \beta^i} $. Below we will show that $ \beta^{a} \equiv (\beta^{0},\beta^{i})$ are the components of the four-velocity of comoving observers in a locally orthonormal frame. We will called this particular frame \textit{the river frame}, the reason is given in the next subsection. The river frame is fully specified by the set of \textit{dual vectors} $ \eb{a} \equiv \lb \eb{0}, \eb{j} \rb $, %
\begin{align}
	\eb{0} = (a e^{ \Phi}) \beta^0\dd \eta \,, \qquad \eb{j} = (a e^{ \Phi}) \lc \lp e^{ - M} \rp ^{j}_{\ i} \dd x^i + \beta^j \dd \eta \rc \,,
	\label{eq:dual-tetrad-basis}
\end{align}
which%
\footnote{Note the presence of the conformal factor $ a e^{\Phi}$. This is needed in order to get quantities in the physical spacetime metric $ \dd s^{2}$.
} %
form an orthonormal (dual-) tetrad basis for the spacetime. In the river frame (as in any orthonormal tetrad frame) the metric looks locally flat, that is, $ \dd s^{2} = \eta_{\underline{ab}}\ \eb{a}\ \eb{b}$, with $ \eta_{ \underline{ab}}$ the Minkowski metric. From the previous equation we see that the dual coordinate basis $ \dd x^{\mu} $ (which we called the \textit{background frame}) is related to the river frame through $ \eb{a} = (a e^{ \Phi}) \e{a}{\mu}\ \dd x^{ \mu}$, where the (conformal) change of basis matrix is 
\begin{align}
	\e{0}{ \mu} = \beta^0 \delta_{ \mu 0} \,,   \qquad  
	\e{i}{0} = \beta^i \,,  \qquad \e{i}{j} = \lp e^{-M} \rp ^{i}_{\ j}\,.
	\label{eq:tetrads}
\end{align}
From this follows that $ g_{\mu \nu} = \e{a}{\mu} \e{b}{\nu}\ \eta_{ \underline{ab}}$. Note that we are using an ``underline'' to distinguish between the components of an arbitrary tensor in the river frame (tetrad components) from the components in the background frame (coordinate components). For instance, for an arbitrary dual vector $ k$, we have $ k = k_{a} \dd x^{a} = \unI{k}{b}\ \eb{b}$. Additionally, tetrad indices are raised and lowered with the flat metric $ \eta_{ \underline{ab} }$ ($ \un{v}{a} = \eta^{ \underline{ab}}\ \unI{v}{b} $ and $ \unI{v}{a} = \eta_{ \underline{ab}}\ \un{v}{b}$) while coordinate components are raised and lowered with the metric $ g_{ab}$. From the previous discussion, it follows that the tetrad components of any four-vector $ k$ can be obtained from the coordinate components by $ \un{k}{b} = (a e^{ \Phi})\, \e{b}{a}\, k^{a}$.

\subsection{The river moves through the background}

Below we justify the name for the river frame. For that, it will be useful to introduce the following concepts: we say that an observer is \textit{comoving with the background} (or simply comoving) if its four-velocity $ \com{u}$ satisfies $ \com{u}^{i} = 0$. Additionally, an observer is \textit{comoving with the river} (or tetrad-comoving) if its four-velocity $ \tilde{u}$ satisfies $ \un{ \tilde{u}}{i} = 0 $. This allow us to give a nice interpretation of the metric components $ \beta^{a}$ as the tetrad components of the four-velocity of comoving observers. Indeed, the four-velocity of comoving observers can be written in the river frame as
\begin{align}
	\un{\com{u}}{a}\  = ( a e^{\Phi})\e{a}{0}\ \com{u}^{0} = 
	\e{a}{0} = \beta^{a}\,.
	\label{eq:u-com-tetrad-frame}
\end{align}
Here we used the normalization condition $u \cdot u = - 1$ to obtain $ (a e^{\Phi})\, \com{u}^{0} = 1$. Given an observer with four-velocity $ u$, its peculiar velocity is defined according to
\begin{align}
	\com{u} = \gamma \lp u - v \rp \,, \qquad \gamma  = - 
	u \cdot \com{u} = \f{1}{ \sqrt{1 - v \cdot v }}\,,
	\label{eq:uv}
\end{align}
where $ v$ belongs to the rest frame of the observer $ u$, that is, $ u \cdot v = 0$. Therefore, we see that $ - v$ is the velocity of comoving-observers w.r.t $ u$. In particular, using $ \un{ \tilde{u}}{a} = (1,0,0,0)$ and  \eq{eq:u-com-tetrad-frame} we obtain \textit{the peculiar velocity of the river} $ \tilde{v} = - \tb$, where
\begin{align}
	\un{ \tb}{a} \equiv \lp 0, \beta^{i}/ \beta^{0} \rp \,, 
	\qquad \textrm{and} \qquad 
	\tilde{\gamma} = \beta^{0} = \lp 1 - \tb \cdot \tb \rp^{- 1/2} \,. 
	\label{eq:vtoCF}
\end{align}
Note that $ \tb $ is the relative velocity of comoving observers w.r.t the river. By using this definition we can rewrite \eq{eq:Cmetric} in a very intuitive way
\begin{align}
	\dd s^2 = - \dd t \sub{R}^{2} + \eta_{\underline{ij}} \lp \un{ \dd x\sub{B}}{i} - \un{ \tilde{v} }{i}\ \dd t\sub{R}\rp 
	\lp \un{ \dd x\sub{B}}{j} - \un{ \tilde{v}}{j}\ \dd t\sub{R}\rp  \,,
	\label{eq:river-metric}
\end{align}
where we introduced the \textit{river's proper time} $ \dd t\sub{R} \equiv (a e^{ \Phi})\, \beta^0\, \dd \eta$, and we have abbreviated $ \un{dx\sub{B}}{j} \equiv (a e^{ \Phi}) \lp e^{-M} \rp^{j}_{\ i} \ \dd x^i$. Do not confuse $ \un{ \dd x\sub{B}}{j}$ with the \textit{river's proper distance}
\begin{align}
	\un{ \dd l\sub{R}}{j} = \un{ \dd x\sub{B}}{i} - \un{ \tilde{v}}{i}\ \dd t\sub{R} \,.
	\label{eq:proper-distance}
\end{align}
\eq{eq:river-metric} give us the picture of a river of space. Indeed, consider an object which is comoving with the river, then along its world-line we have $ \un{ \dd l\sub{R}}{j} = 0$, that is, there is no displacement w.r.t the river. However, in a proper time $ \dd t\sub{R}$ the object displaces by $ \un{ \dd x\sub{B}}{i} = \un{ \tilde{v} }{i}\ \dd t\sub{R}$ w.r.t the background (w.r.t comoving observers). To be precise, 
from the point of view of the river, it is the background which is moving, so that in a proper time $ \dd t\sub{R}$ the background displaces by $ \un{ \dd x\sub{B}}{i} = - \un{ \tb }{i}\ \dd t\sub{R}$.

\subsection{Fishes in the river}

%
%
\begin{figure}[h!]\centering
	\includegraphics[scale=0.5]{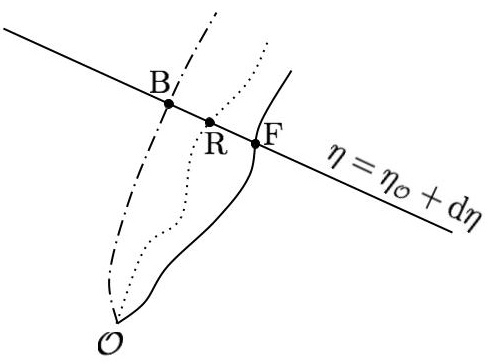}
	\caption{A fish, a tetrad-comoving-object (comoving with the river) and a comoving object (comoving with the background) start together in $ {\mathcal O}$ at conformal time $ \eta$. At time $ \eta + \dd \eta$ they will be at F,R and B respectively. Their relative separation satisfies $ d \sub{ \textrm{F,B}} =  d \sub{ \textrm{F,R}} +  d \sub{ \textrm{R,B}}$.}\label{fig:river}
\end{figure}

Consider a fish swimming in the river. According to the river, in an proper time $ \dd t\sub{R}$ the fish moves a proper distance $ \un{ \dd l\sub{R}}{i}$. In the same interval of time and from the point of view of the river, the fish get apart from the background by
\begin{align}
	\un{ \dd x\sub{B}}{j} = \un{ \dd l\sub{R}}{i} + \un{ \tilde{v}}{i}\ \dd t\sub{R} \,.
\end{align}
In figure~\ref{fig:river} we provide a schematic representation for the path followed by comoving-objects, river-comoving-objects and fishes. If at conformal time $ \eta \sub{ {\mathcal O}}$ the three types of objects are at the same  spacetime point $ {\mathcal O}$, they will move apart as time goes on. At conformal time $ \eta \sub{ {\mathcal O}} + \dd \eta$ they will be at positions B,R and F respectively. According to the river, the distance between the fish and the background ($ d \sub{ \textrm{F,B}}$) is equal to the distance between the fish and the river ($ d \sub{ \textrm{F,R}}$) plus the distance between the river and the background ($ d \sub{ \textrm{R,B}}$), that is, $ d \sub{ \textrm{F,B}} =  d \sub{ \textrm{F,R}} +  d \sub{ \textrm{R,B}}$.

According to the river, the fish has a velocity 
\begin{align}
	\un{ \vf}{i} = \f{ \un{ \dd l\sub{R}}{i}}{ \dd t\sub{R}} = 
	\f{ 1}{\beta^{0}} \lp e^{-M} \rp^{i}_{\ j} \ \der{x^j}{ \eta} + \un{ \tb}{i}\,.
	\label{eq:vF}
\end{align}
This result is valid for any fish, even for photons. In the particular case of photons, $ \dd s^{2} = 0$, and it follows from \eq{eq:river-metric} that $ \un{ \vf}{i}\ \unI{( \vf )}{i} = 1$, that is, photons move with speed $ c \equiv | \vf| = 1$ w.r.t the river. Let's obtain these results in a different way.

In general, the four-velocity $ u$ of any observer (any fish) can be written as $ u = \tilde{\gamma}  \sub{F}  \lp \tilde{u} + \vf  \rp $, and using $ \un{ \tilde{u}}{a} = (1,0,0,0)$ we obtain
\begin{align}
	\tilde{\gamma}  \sub{F} = \un{u}{0} \,,  
	\qquad \un{ \vf }{a}   = \lp 0, \un{u}{i}/ \un{u}{0} \rp\,.
	\label{eq:vtoCF}
\end{align}
Then expressing the tetrad components in terms of the coordinates components, $ \un{u}{b} = (a e^{ \Phi}) \e{b}{a}\ u^{a}$, and using \eq{eq:tetrads} we obtain
\begin{align}
	\f{ \un{u}{i}}{ \un{u}{0}} = \f{ 1}{\beta^{0}} 
	\lp e^{-M} \rp^{i}_{\ j} \ \f{u^{i}}{ u^{0}}+ \un{ \tb}{i}\,.
	\label{eq:ui-u0}
\end{align}
Since $ u^{a} = u^{0}(1, \dd x^i/ \dd \eta )$ we see that this equation coincides with \eq{eq:vF}. For photons the previous equation is valid if we change the four-velocity $ u^{a}$ by the photon's four-momentum $ p^{a}$.

We remind the reader that for an observer $ u$, the four-momentum $ p$ of a given a photon can be written as
\begin{align}
	p = E \lp u - n \rp \,, \qquad  \textrm{with}
	\qquad u \cdot n = 0 \,, \qquad E = - p \cdot u\,, 
	\label{eq:n}
\end{align}
where $ E$ and  $ n^{a} $ are respectively, the observed energy and direction of arrival in the $ u$-frame. In particular, in the river frame  the energy and direction of incoming photons has a simple form
\begin{align}
	\tilde{E}  & = - \unI{p}{0} \,, \qquad  \unI{ \tn}{a}  = \lp 0, \unI{p}{i}/ \unI{p}{0} \rp \,.
	\label{eq:EtoCF}
\end{align}
Note also that, since the velocity of photons in the river frame is $ c = 1$, then in that frame the four-velocity of photons  is just $ - \tn$. Therefore, it follows from \eq{eq:vF} that the direction of arrival of photons (in the river frame) satisfies
\begin{align}
	 - \un{\tn}{i} = 
	\f{ 1}{\beta^{0}} \lp e^{-M} \rp^{i}_{\ j} \ \der{x^j}{ \eta} + \un{ \tb}{i}\,.
	\label{eq:tn}
\end{align}
Further relations can be obtained from \cite{Roldan:2017wvm}. We now quickly compare our results with the existing river model for black holes.

\subsection{Comparison with the Gullstrand-Painlev\'e metric}
\label{sec:river-black holes}

Although \eq{eq:river-metric} have been written in a way that mimics the Gullstrand-Painlev\'e Cartesian metric (see the introduction of this paper)
\begin{align}
		\dd s^2 = - \dd t_{ff}^{2} + \eta_{ij} \lp \dd x^{i} - \beta^{i}_{r}\ \dd t_{ff}\rp 
	\lp \dd x^{j} - \beta^{j}_{r}\ \dd t_{ff}\rp  \,,
	\label{eq:gull-metric}
\end{align}
there are however significant differences as we stress below. Note that the Gullstrand-Painlev\'e metric describes a particular geometry (the Schwarzschild one) and is given in a  fixed coordinate system $ x^{\mu} = ( t_{ff},x^{i})$ and so the quantities $ \dd t_{ff}$ and $ \dd x^{i}$ are truly differentials. By contrast, in our case we have no prescribed any geometry in the sense that the $ \Phi$, $ \beta^{i}$ and $ M_{ij}$ are arbitrary fields. For this reason, \eq{eq:river-metric} should be valid for a wide range of curved spacetimes, mainly we think it can be useful in cosmological models and the large-scale structure. For instance, two metrics of cosmological interest, Eqs.(6) and (14) of \cite{Mizony:2004sh}, can immediately be written in our river version of a metric.

On the other hand, in \eq{eq:river-metric} there is no fixation of the coordinates. Indeed, let's think of 
$ \Phi$, $ \beta^{i}$ and $ M_{ij}$ as perturbations to the FLRW spacetime (see \eq{eq:Cmetric}). The fact that the coordinate system is still arbitrary follows from the well-known freedom of cosmological perturbations on the FLRW spacetime, where we can change the coordinates $ x^{\mu} = ( \eta, x^{i})$ to a new one by the transformation 
\begin{align}
	\eta \to \eta_{*} = \eta - \alpha \,, \qquad x^{i}_{*}= x^{i} - \xi^{i} \,,
\end{align}
with infinitesimal parameters $ \xi^{\mu} = ( \alpha, \xi^{i})$. Because of this generality, we see that (in general) neither $ \dd t\sub{R}$ nor $ \un{ \dd x\sub{B}}{j}$ are truly differentials. That is, $ \dd (\dd t\sub{R}) \ne 0$. Indeed, from the definitions of $ \dd t\sub{R}$ and $ \un{ \dd l\sub{R}}{j}$ we see that they are the one-forms $ \eb{a} \equiv \lb \eb{0}, \eb{j} \rb $ given in \eq{eq:dual-tetrad-basis}.

Note however that $ \dd t\sub{R}$ is a truly differential along the world-line of tetrad-comoving-observers (observers comoving with the river). Indeed, for one such observer with four-velocity $ \tilde{u}$ we have
$ \dd t\sub{R} = \tilde{u}_{\mu} \dd x^{\mu}$. Because of the path dependence of these differentials, we can think of $ \dd t\sub{R}$ and $ \un{ \dd x\sub{B}}{j}$ as \textit{inexact differentials} in a similar way as is given in thermodynamics for the heat and work. Despite this fact, note that results we obtained by using the inexact differentials (e.g., \eq{eq:vF}) are the same we can obtain by using a more formal treatment (as we did in Eqs.~\eqref{eq:vtoCF}-\eqref{eq:ui-u0}). Additionally, the analogy of the river is useful because it provides an intuitive view of the physics we are discussing.

\section{CMB up to second order}
\label{sec:CMB2nd}

In this section we write explicitly the second-order Sachs-Wolfe formula, and in \refs{sec:GInvariance} we proof its gauge invariance. From Eqs.~\eqref{eq:Gsw-intro}, \eqref{eq:I0} and \eqref{eq:Doppler-river}, the (logarithmic) temperature anisotropies up to second order are given by 
\begin{align}
	\Theta  = \lc \cT + \Phi + (\tb - \vf) \cdot \tn + 
	\f{ \tb \cdot \tb_\perp - \vf \cdot \vf^{ \perp} }{2} \rc \Big|_{ \eta_o}^{\eta_e} +
	\int_{ \eta_e}^{ \eta_o} \dd\eta  \lp \tb' + 
	\tn \cdot M' \rp \cdot \lp \tn + \tb_\perp \rp \,,
	\label{eq:sw2nd-app}
\end{align}
where $\tb_\perp$ is the orthogonal projection of $ \tb$ on $ \tn$, that is, $ \tb_\perp = \tb - \tn ( \tn \cdot \tb)$, and similarly for $ \vf^{\perp}$. We will find it convenient to define $ \tilde{V}  \equiv \vf - \tb$. Then note that 
\begin{align}
	\tb \cdot \tb_\perp - \vf \cdot \vf^{ \perp}  & = 
	(\tb - \vf) \cdot (\tb + \vf) - \lc (\tb - \vf) \cdot \tn \rc \lc (\tb + \vf) \cdot \tn \rc \notag \\
	 & =  - \tilde{V}  \cdot \lp 2\tb + \tilde{V}  \rp_{\perp} \,,
\end{align}
so the \textit{second-order Sachs-Wolfe formula} can be written as%
\begin{align}
	\boxed{\Theta = \lc \cT + \Phi - \tilde{V}  \cdot \tilde{N}\rc \Big|_{ \eta_o}^{\eta_e} +
	\int_{ \eta_e}^{ \eta_o} \dd\eta  \lp \tb' + 
	\tn \cdot M' \rp \cdot \lp \tn + \tb_\perp \rp \,, }
	\label{eq:sw2nd-curved}
\end{align}
and we defined the four-vector $ \tilde{N}\equiv \tn + \tb_{\perp} + \tilde{V} _{\perp}/2$. We want to stress that $ \tilde{V} , \tilde{N}, \tb$ are four-vectors belonging to the rest space of the river, because of that, their $ 0-$component in the tetrad-frame vanish. That is, $ \un{\tilde{V} }{0} = \un{ \tilde{N}}{0} = \un{ \tb}{0} = 0$.

We remind the reader that the peculiar velocity, $ v$, is defined by the relation $ \com{u} = \gamma \lp u - v \rp$, which up to second order gives $ \un{v}{i} = \un{u}{i} - \un{ \tb}{i}$. Additionally, from \eq{eq:vtoCF} we know that the fish's velocity is $ \un{ \vf }{a}   = \lp 0, \un{u}{i}/ \un{u}{0} \rp$. From this see that up to second order
\begin{align}
	\un{v}{i} = \un{ \tilde{V}}{i}\,,
	\label{eq:Vi}
\end{align}
that is, $ V$ has the same $ i-$tetrad components that the peculiar velocity. Note however that $ \un{v}{0} \ne 0$, which is a consequence of the fact that $ v$ belongs to the rest frame of $ u$%
\footnote{Indeed, $ -v$ is the velocity of comoving observers as seen by $ u$.
} %
not to the rest frame of the river (as is the case of $ \tilde{V}$).

To prove the gauge invariance of the second-order $ \Theta$ we do not need the explicit form of $ \tn$ or the photon's (curved) path $ x^{i}( \eta)$. However, we will provide these relations just for completeness. These quantities where obtained in \cite{Roldan:2017wvm}. At any point along the photon's trajectory, the direction vector is given by
\begin{align}
	\un{\tn}{i}( \eta) = \un{\tn}{i}_{o} - 
	\lc \un{\lp M \cdot \tn \rp}{i} - 
	\un{\tn}{i}\ \lp \tn \cdot M \cdot \tn \rp \rc \Big|_{\eta}^{\eta_o} - 
	\int_{ \eta}^{ \eta_o} \dd \eta\ \p{i}^{\perp} \lp \tn \cdot \tb + 
	\tn \cdot M \cdot \tn \rp \,.
	\label{eq:ni-1st-sol}
\end{align}
Note that we just need the direction vector $ \tn$ up to first order, this is so because in \eq{eq:sw2nd-curved} $ \tn$ is always multiplying quantities which are at least first order. For the same reason, on the right hand side of the previous equation we can simply set $ \tn$ at zero order, that is, $ \un{\tn}{i} = \un{\tn}{i}_{o}$. In \eq{eq:sw2nd-curved} the integration must be performed along the photon's (curved) path, whose coordinates are given by 
\begin{align}
	x^{i}( \eta)  & = x^{i}_{o} +  
	\lc \un{\tn}{i} - \un{\lp M \cdot \tn \rp}{i} + 
	\un{\tn}{i}\ \lp \tn \cdot M \cdot \tn \rp \rc \Big|^{\eta_o} \lp \eta_o - \eta \rp
	\label{eq:xi-1st-sol}\\
 	& + \int_{ \eta}^{ \eta_o} \dd \bar{\eta}\ \lc \tb^i + 2 \un{\lp M \cdot \tn \rp}{i} - 
	\un{\tn}{i}\ 	\lp \tn \cdot M \cdot \tn \rp \rc - 
	\int_{ \eta}^{ \eta_o} \dd \bar{\eta}\ \lp \bar{ \eta} - \eta \rp \p{i}^{\perp} 
	\lp \tn \cdot \tb + \tn \cdot M \cdot \tn \rp \,. 
	\notag
\end{align}
\OLD{\eq{eq:sw2nd-curved} can also be written in an alternative form, using the so called Born approximation. See section~4 of \cite{Roldan:2017wvm} for details. 
}

Eqs.~\eqref{eq:ni-1st-sol} and \eqref{eq:xi-1st-sol} are not independent. Indeed, if $ p^{a} $ is the photon's four-momentum, then we define $ q^{a}$ (which is not a four-vector) by the relation
\begin{align}
	q^{a} \equiv \f{ p^{a}}{p^{0}}  \,, \quad \Rightarrow  \quad 
	q^{a} = \der{x^{a}}{\eta} = \lp 1, \dot{x}^{i} \rp \,.
	\label{eq:qDEF}
\end{align}
Now, by using \eq{eq:tn} we obtain up to first order
\begin{align}
	\un{\tn}{i} = - \lp q^{i} + \tb^i + M_{ij}\ \un{\tn}{j} \rp \,.
\end{align}
The coordinates of the photon's trajectory are just $ x^{a}( \eta)  = \int \dd \eta\ q^{a}$, and we also have the useful relation $ \dd/ \dd \eta = q^{c} \p{c}  = \p{0} + q^{i} \p{i}$. The latter relation can alternatively be written as
\begin{align}
	\der{}{ \eta} = \p{0} - \un{ \lp \tn + \tb + M \cdot \tn \rp }{i}\ \p{i} \,,
	\label{eq:deta}
\end{align}
from which we get at zero order: $ \dd/ \dd \eta = \p{0} - \un{\tn}{i}\, \p{i}$. We are now ready to study how gauge transformations affect the relevant quantities entering the CMB temperature anisotropies $ \Theta$. This will be the focus of the next sections. In particular, in \refs{sec:GInvariance} we show the gauge invariance of the second-order Sachs-Wolfe formula \eq{eq:sw2nd-curved}.

\section{GT (gauge transformations) up to second order}
\label{sec:gt2nd}

In this section we review the concept of gauge transformations up to second order. We introduce a notation which will facilitate our calculations and provide a set of formulas which we think can be used as a reference for other works involving gauge transformations at second order.

Given a geometrical object $ T$ (it can be: a scalar, a vector, a tensor, the Christoffel symbols, etc.), and a vector field $ \xi^{\mu} = (\alpha,\xi^{i})$, the gauge transformation of $ T$ induced by $ \xi^{\mu}$ is defined by \cite{Bruni:1996im,Matarrese:1997ay,Malik:2008im}
\begin{align}
	T_{*} = e^{ \lie} T  =  T + \lie T + \f12 \lie (\lie T) + \cdots\,,
\end{align}
where $ \lie$ is the Lie derivative along $ \xi^{\mu}$. Up to second order in $ \xi^{\mu}$, this expression can be written in a nice form
\begin{align}
	T_{*} = T + \f12 \lie \lp T + T_{*} \rp \,,
	\label{eq:gt-rule1}
\end{align}
which easily allow us to compute the second-order gauge transformation if we already know the first-order one. If additionally we write $ T_{*} = T + \Delta T$, the equation above can equivalently be written as
\begin{align}
	\Delta T = \lie \lp T + \f{\Delta T}{2} \rp \,.
	\label{eq:gt-rule2}
\end{align}
In this work we will use these two notations extensively. The Lie derivative acting on scalar $ S$, vector $ V^{a}$ and tensor $ T_{ab}$ are given by \cite{Carroll:1997ar}
\begin{align}
	\lie S = \xi^{c} \p{c}\, S \,, \quad \lie V^{a} = \xi^{c} \p{c} V^{a}- V^{c} \p{c}\, \xi^{a} \,, \quad 
	\lie T_{ab} = \xi^{c} \p{c} T_{ab} + \lp T_{ac} \p{b} + T_{cb} \p{a} \rp \xi^{c}\,.
	\label{eq:lieDer}
\end{align}
These relations will be useful below. Note that in the previous equations we have used the coordinate components, not the tetrad indices.

A gauge transformation is by definition a diffeomorphism which takes an arbitrary point $ P$ (on the spacetime) into another point $Q$ in the same spacetime, and the previous equations show explicitly how some geometrical objects transform under this diffeomorphism.  Gauge transformations are also known as \textit{active transformations}. As we review below, they are closely related to a change of coordinates (passive transformations).

In the active approach there is a map which takes an arbitrary point $ P$ into another point $Q$, but the coordinates of the spacetime are kept fixed (the charts on the manifold are fixed). Naturally, the coordinates of the point $ Q$ are different from those of $ P$, and the relation is given by
\begin{align}
	x^{\mu}(Q) & = e^{ \xi^{c} \p{c}} x^{\mu} (P) = 
	%
	x^{\mu}(P) + \xi^{\mu}(P) + \f12 \xi^{c} \p{c}\, \xi^{\mu} + \cdots 
	\label{eq:xmuP-Q}
\end{align}
In the passive approach, the transformation is made at the same point in the spacetime. That is, a new chart on the manifold is chosen (but there is no remapping of the manifold). To relate the two approaches, we choose the new coordinates $ y^{\mu}$ of the point $ Q$ to be the coordinates $ x^{\mu}$ of the point $ P$, that is
\begin{align}
	y^{\mu}(Q) & \equiv x^{\mu}(P) = e^{ - \xi^{c} \p{c}} x^{\mu} (Q) = 
	%
	x^{\mu}(Q)  - \xi^{\mu} \lp x(Q) \rp + \cdots 
	\label{eq:ymu}
\end{align}
Although the two approaches to gauge transformations are equivalent, in this work we will use the active one (in which the coordinates of the spacetime are kept fixed). One good reason for this is that the definition of the the last scattering surface remains simple even after applying a gauge transformation. Indeed, if we define the last scattering surface $S_{e,o}$ to be placed at the fixed time of emission $ \eta_e$ (see before \eq{eq:cT_o}), then after a GT the coordinates of the spacetime remains the same and so $S_{e,o}$ still belongs to the hypersurface of $ \eta = \eta_e = const$. This is not the case however if we use a coordinate transformation in which $ \eta \to \hat{ \eta} (\eta, x^{i})$ (see \cite{Mirbabayi:2014hda} for an specific example).

\OLD{ motivation for this is the following: as we discussed in \refs{sec:cT}, for $ \eta \le \eta_e$ the temperature of the photon-baryon fluid is written as $ T = \braket{ T}\, e^{ \cT}$ where $ \braket{T} \propto 1/a(\eta)$ is the background temperature. In this sense,  the hypersurface of $ \eta = const$ is defined by the physical argument that $ \braket{T} = const$ as in Mirbabayi \& Zaldarriaga \cite{Mirbabayi:2014hda}. So in the sudden decoupling approximation, CMB photons were emitted at a fixed time $ \eta_e$ when the mean temperature $ \braket{T}_{e}$ corresponds roughly to the ionization energy of the hydrogen. Because of this, when transforming the CMB temperature, it is better to use active transformations (acting on the fields) rather than passive transformations (acting on the coordinates).
}

\OLD{, because in the latter case the transformation of the time coordinate becomes intricate.
}

Although in the active approach the coordinates of the spacetime are fixed, this is not the case for the coordinates $ x^{i}( \eta)$ of the photon's geodesic. The reason is simple: gauge transformations act on the metric field, and the photon's coordinates depend on the metric (see \eq{eq:xi-1st-sol}). This fact turns out to be very relevant and makes the gauge transformations of the CMB non trivial.

\OLD{Changing the coordinates of the spacetime is however different from changing the coordinates of the photon's geodesic. This is because the coordinates of the spacetime are totally independent on the metric fields, while the coordinates of the photon's geodesic do depend. This will be shown explicitly in the next section.
}

In the next subsections we provide the gauge transformations of metric and fluid fields, as well as the transformations of the photons variables $ x^{i}( \eta)$ and $ \un{\tn}{i}$. Then in \refs{sec:GInvariance} we prove the gauge invariance of the second-order Sachs-Wolfe formula \eq{eq:sw2nd-curved}.
\OLD{ which describes the CMB temperature anisotropies up to second order.}

\subsection{GT of the metric and fluid perturbations}
\label{sec:GT-metric}

The transformation rules for the metric and fluid perturbations are derived in the Appendix~\ref{app:2ndGT}. Here we collect all the relevant expressions that will be used to prove the gauge invariance of $ \Theta$. As discussed before, for an arbitrary tensor $ T$, we write the transformed one as $ T_{*} = T + \Delta T$. Below we provide the transformation for the metric
\begin{align*}
	\dd s^2  &  = a^2(\eta) e^{2\Phi} \lc - \dd \eta^2 + 
	2 \beta^{j} \lp e^{-M} \rp_{ji} \ \dd x^i \dd \eta 
	+ \lp  e^{-2M}\rp_{ij}\dd x^i \dd x^j \rc \,.
\end{align*}
Since $ \un{ \tb}{a}  = \lp 0, \beta^{i}/\beta^{0} \rp$, we have that up to second order $ \un{ \tb}{i} = \beta^{i}$. We will also provide the gauge transformation of the $ i-$tetrad components of peculiar velocity $ v$, and as we saw in \eq{eq:Vi} up to second order we have $ \un{v}{i} = \un{ \tilde{V}}{i}$. So taking into account these facts, the results of this section give us the necessary tools to prove the gauge invariance of $ \Theta$.

\paragraph{Metric perturbations}
%
%
\begin{align}
	\Delta \Phi & = \lc \alpha' +\cH \alpha \rc  + 
	 \xi^{c} \p{c} \lp \Phi + \Delta \Phi/2 \rp -  \lp \beta + \Delta \beta/2 \rp_i \xi_{i}'\,,
	\label{eq:GT-phi} \\
	\Delta \beta_{i} & = \lc - \alpha_{,i} + \xi_{i}' \rc +  
	 \xi^{c} \p{c} \lp \beta + \Delta \beta/2 \rp_{i} + 
	\f12 \p{[i}\ \xi_{j]} \lp \beta + \Delta \beta/2 \rp_{j} - 
	 \lp \alpha_{,j} + \xi_{j}'\rp \lp M + \Delta M/2 \rp_{ji}\,,
	\label{eq:GT-B}
\end{align}
\begin{align}
	\Delta M_{ij} =  & \lc \alpha'\ \delta_{ij} - \xi_{(i,j)} \rc +    
	 \xi^{c} \p{c} \lp M + \Delta M/2 \rp_{ij} 
	 -  \lp \beta + \Delta \beta/2 \rp_{k} \xi_{k}'\ \delta_{ij} 
	\notag\\
	 & + \f12\lb \p{[i}\ \xi_{k]}\ \lp M + \Delta M/2 \rp_{kj} 
	- \lp M + \Delta M/2 \rp_{ik} \p{[k}\ \xi_{j]} \rb
	-  \lp \beta + \Delta \beta/2 \rp_{(i} \p{j)} \alpha \,,
	\label{eq:GT-M}
\end{align}
where $ \xi^{\mu} = (\alpha,\xi^{i})$ is the vector field generating the gauge transformation. The notation $ (,)$ means symmetrization and $ [,]$ anti-symmetrization, that is, $ \xi_{(i,j)} \equiv (\p{i}\xi_{k} + \p{k}\xi_{i})/2$ and $ \p{[i}\ \xi_{k]} \equiv \p{i}\xi_{k} - \p{k}\xi_{i}$. Note also that $ \alpha_{,i} \equiv \p{i} \alpha$.

To get compact expressions, in this work we will adopt the following notation: indices in $ \xi^{i}$ are raised and lowered with a delta Kronecker,%
\footnote{This is a convenient convention. We will never need to use the covariant vector $ \xi_{a} = g_{ab} \xi^{b} $, so no confusion should arise.
} %
that is $ \xi^{i} = \xi_{i}$. Because of that, we can think of $ \xi^{i}$ as inducing a vector field in the local orthonormal frame (the river frame), with tetrad indices given by $ \un{ \xi}{a} \equiv (0, \xi^{i})$. This notation is nice because we can use scalar products between $ \xi$ and other quantities like $ \tb$, $ M$, and $ \tilde{V}$ which we always think of as being represented in the river frame. So for instance, in the next sections we will extensively use the notation $ \tn \cdot \xi \equiv \un{\tn}{i}\, \xi_{i}$, $ \tb \cdot \xi \equiv \tb^{i} \xi_{i}$, and so on.

\paragraph{Fluid perturbations}
%
%
The (logarithmic) intrinsic temperature perturbation and the peculiar velocity transform as
\begin{align}
	\Delta \cT & = - \cH \alpha +  \xi^{c} \p{c} \lp \cT + \Delta \cT/2 \rp \,, 
	\label{eq:GT-cT}\\
	\Delta \un{v}{i} & = - \xi_{i}' + 
	 \xi^{c} \p{c} \unI{ \lp v + \Delta v/2 \rp }{i} + 
	\f12 \p{[i}\ \xi_{j]} \unI{ \lp v + \Delta v/2 \rp }{j} + 
	 \lp  M + \Delta M/2 \rp_{ij}  \xi_{j}' \,.
\end{align}
Note that the first-order relations are simply obtained by neglecting the explicitly quadratic terms. For instance, at first order we have: $ \cT_{*} =  \cT - \cH \alpha$. Additionally, by using this first-order relation we can rewrite \eq{eq:GT-cT} as $ \cT_{*} =  \cT - \cH \alpha + \xi^{c} \p{c} \lp \cT- \cH \alpha/2 \rp$, and analogously for the others relations. So the expressions we give above seems useful for recursive computation. 

\OLD{For instance, suppose that you transform to the synchronous gauge in which $ \beta^{i}_{*} = 0$, then Eqs.~\eqref{eq:}-\eqref{eq:} simplify a somehow (as opposite to the existing versions in literature which are totally expressed in terms of $ \xi^{a}$).}

Note the presence of the factor $ \xi^{c} \p{c}$ in all the gauge transformations given above. This can understood as follows: intuitively we can understand $ \Delta T$ as being the difference between the tensor field $ T$ at the spacetime point $ Q$ with the lie drag of $ T$ from the spacetime point $ P$ to $ Q$. 
Then the factor $ \xi^{c} \p{c}$ takes into account the fact that $ Q$ and $ P$ have different coordinates $ x^{\mu}(Q) = x^{\mu}(P) + \xi^{\mu}(P) + \cdots$, see \eq{eq:xmuP-Q}. In other words, if we write $ \Delta x = x^{\mu}(Q) = x^{\mu}(P)$

This observation will be important below.

Finally, note that in principle the intrinsic temperature anisotropies $ \cT$ depend not only on position $ x$ but also on direction $ \tn$. So in the case of $ \cT$, the notation $ \xi^{c} \p{c}$ must be understood as a compact notation for the change in position and also for the change in direction. In the next section we show how the direction vector $ \tn$ changes under a gauge transformation.

\subsection{GT of the photon's variables}
\label{sec:GT-nixi}

Hereafter we focus on the subtle issue of (second-order) gauge transformations as applied to the CMB anisotropies, an issue that was discussed in \cite{Mirbabayi:2014hda} although from a different point of view (by using passive transformations) and considering a particular type of transformations. Here however we will be fully general. Below, we firstly compute the gauge transformations of the photon's four-momentum $ p^{a}$ and then by using the relation $ q^{a} = p^{a}/p^{0}$ we will find the transformation rules for $ q^{a}$ and consequently for $ \un{\tn}{i}$ and $ x^{i}$. Note that for the second-order temperature anisotropies we only need $ q^{a}$ up to first order.

\subsection*{GT of the photon's path coordinates}

Up to first order, the gauge transformation of the photon's four-momentum is given by
\begin{align}
	p^{a}_{*}  & = p^{a} + \lp \xi^{c} \p{c}\, p^{a} - p^{c} \p{c}\, \xi^{a} \rp 
	= p^{0} \lc q^{a} - \lp 2 \cH \alpha\, q^{a} + \dot{ \xi}^{a} \rp  \rc  \,,
\end{align}
where we have used \eq{eq:lieDer}, the fact that $ q^{c} \p{c} = \dd/ \dd \eta$, and that at zero order $ q^{a} = const$ while $ p^{0} \propto 1/a^{2}$. Dividing by $ p^{0}_{*}$ on both sides of the previous equation yields 
\begin{align}
	q^{a}_{*} = q^{a} + \dot{ \alpha}q^{a} - \dot{ \xi}^{a} \,, 
	\qquad \to \qquad 
	q^{i}_{*} = q^{i} - \lp \dot{ \alpha}\, \un{\tn}{i}  + \dot{ \xi}^{i} \rp \,,
\end{align}
where we used that at zero order $ \un{\tn}{i} =  - q^{i}$. Note also that the following relation holds, $ q^{0}_{*} = q^{0} = 1$. Then remembering that the trajectory of the photon has coordinates $ x^{a}( \eta)  = \int \dd \eta\, q^{a}$, we get after integration
%
%
\begin{align}
	\eta_{*} = \eta \,, \qquad x^{i}_{*} = x^{i} - \lp \alpha\, \un{\tn}{i} + \xi^{i} \rp  \,.
	\label{eq:x*}
\end{align}
Compare this result with the naive transformation of coordinates $ x^{a}_{*} = x^{a} - \xi^{a}$ that we would expect from \eq{eq:ymu}. Then according to the discussion of the previous section, we need to use the rule%
\footnote{This is \texttt{Rule 1}, a second rule will be given in \refs{sec:GTisw}.
} %
\begin{align}
	\texttt{Rule 1:} \qquad \qquad 
	\xi^{c} \p{c} \to \lp \alpha\, \un{\tn}{i} + \xi^{i} \rp \p{i} \,,
	\label{eq:rule1}
\end{align}
that is, the time derivative $ \p{0}$ is changed by a radial derivative,%
\footnote{Geodesic as parametrized by affine parameter would transform as expected but as parametrized by constant time there is a need to convert time displacement to radial displacement.
} %
 $ \p{r} = \un{\tn}{i}\, \p{i}$. Let's discuss a bit on this. In the Born approximation,%
\footnote{That is, neglecting the effect of the field perturbations on the photon's path.
} %
 the photon's trajectory is given by $ x^{i} = x^{i}_{o} +  \un{\tn}{i}_{o} \lp \eta_o - \eta \rp$, so at conformal time $ \eta_{e}$ the photon is at a distance $ r_{e} = \eta_o - \eta_{e}$ from the observer. When including the metric perturbations we obtain from \eq{eq:xi-1st-sol}
\begin{align}
	r_{e} \equiv \lp x^{i}_{e} - x^{i}_{o} \rp \un{\tn}{i}_{o} = 
	\eta_o - \eta_{e} + \int_{ \eta_{e}}^{ \eta_o} \dd \eta
	\lp \tn \cdot \beta + \tn \cdot M \cdot \tn \rp \,,
	\label{eq:TDelay}
\end{align}
where a summation on the repeated indices is understood. The previous relation gives the so called time-delay. This equation can be interpreted in two ways. First, integrating to a fixed $ \eta_{e}$ (as we explicitly did) gives a radial displacement, showing that the true emission points are not located in an spherical shell of radius $ r_{e} = \eta_o - \eta_{e}$, but belong to a distorted surface whose ``radius'' depends on direction $ \tn$. Alternatively, integrating to a fixed $ r_{e}$ implies a variation in the conformal time at emission (one such version is given for instance in Eq.7 of \cite{Yoo:2009au}). It is common to define decoupling to happen at a fixed constant time, and this is well motivated by the physical condition $ \braket{T} \propto 1/a( \eta)$. So, although in principle a gauge transformation induces a coordinate transformation given by $ x^{a}_{*} = x^{a} - \xi^{a}$, the condition that decoupling happens at a fixed time $ \eta$ means that the time shift $ \eta_* = \eta - \alpha$ must be changed for a radial displacement. Consequently, the time derivative is changed by a radial derivative as given in \eq{eq:rule1}.

Physically, \texttt{Rule 1} is well supported by the previous discussion and specially by the paragraph after \eq{eq:X*}). However, at the moment I have no rigorous proof of how this rule can be obtained from a mathematical point of view. So in this paper I take \texttt{Rule 1} as a postulate%
\footnote{The same applies to \texttt{Rule 2}, see \eq{eq:rule2}.
} %
and leave their derivation for a future work.

\subsection*{GT of the direction vector}

Using the first-order relation $\un{\tn}{i} = - \lp \beta^i + q^{i} + M_{ij}\, \un{\tn}{j} \rp $, we obtain
\begin{align}
	 \un{\tn}{i}_{*} & =  - \lp \beta_i - \alpha_{,i} + \xi_{i}' \rp - 
	 \lp q^{i}  + \dot{ \alpha}\ q^{i} - \dot{ \xi}^{i} \rp - 
	 \lc M_{ij} + \alpha' \delta_{ij} - \xi_{(i,j)} \rc \un{\tn}{j}\notag \\
	  & = \un{\tn}{i} + \p{i}^{\perp} \alpha + \f12  \p{[i}\ \xi_{j]}\ \un{\tn}{j} \,.
\end{align}
This expression can be rewritten in a compact form as
\begin{align}
	\tn_{*} = \tn + \p{ \perp} \alpha + 
	\f12 \lc \p{} ( \tn \cdot \xi) - (\tn \cdot \p{}) \xi \rc  \,,
	\label{eq:GTni}
\end{align}
where it is understood that in the expression $ \p{} ( \tn \cdot \xi)$, the partial derivative acts only on the field $ \xi$. That is, at zero-order, the direction vector $ \tn$ is treated as a constant. We remind the reader that we are treating $ \xi$ as being a four-vector in the river frame, so $ \tn \cdot \xi = \un{\tn}{i}\ \xi_{i}$. Additionally, to simplify the notation of the sections below, we will understand $ \p{}$ as a shorthand for $ \p{i} $, and so $ \tn \cdot \p{} \equiv \un{\tn}{i}\ \p{i}$. Below, we will use this notation extensively. For instance, we will also use $ \tb \cdot \p{} \equiv \un{ \tb}{i}\, \p{i} = \beta^{i} \p{i}$.

\section{Gauge invariance of {\small $ \Theta = X + \int \dd\eta  \lp Y + Z \rp$}}
\label{sec:GInvariance}

In this section we prove the gauge invariance of second-order Sachs-Wolfe formula, that is, the gauge invariance of $ \Theta$ (see \eq{eq:sw2nd-curved}). We proceed as follows: even though our expressions are fully second order in the metric and fluid fields, we will consider only the gauge transformations at linear order in the gauge parameter $ \xi^{a}$. The reason is the following: if $ \Theta$ is invariant under these gauge transformations, then it will be invariant under any successive gauge transformation, and that proves that it is invariant at any order in $ \xi^{a}$ (see Appendix~\ref{app:quadratic_xi} for more details). Saying that the gauge transformations are linear in $ \xi^{a}$ means for instance, to ignore the terms $ \xi^{c} \p{c} (\Delta \Phi/2)$ and $ (\Delta \beta/2)_i\, \xi_{i}'$ in \eq{eq:GT-phi}.

\OLD{Without loosing generality we will only consider the gauge transformations at linear order in the gauge parameter, $ \xi^{a}$, but which are second-order in the fields $ \Phi, \beta_{i}, \cdots$. If $ \Theta$ is invariant under these gauge transformations, then it will be invariant under any successive gauge transformation, and that proves that it is invariant at any order in $ \xi^{a}$. Saying that the gauge transformations are linear in $ \xi^{a}$ means for instance, to ignore the terms $ \xi^{c} \p{c} (\Delta \Phi/2)$ and $ (\Delta \beta/2)_i \xi_{i}'$ in \eq{eq:GT-phi}. 
}

To proceed, we write the logarithmic temperature anisotropies as
\begin{align}
	\Theta = X\Big|_{ \eta_o}^{\eta_e} + \int_{ \eta_e}^{ \eta_o} \dd\eta  \lp Y + Z \rp \,, 
	\label{eq:ThetaXYZ}
\end{align}
where we defined the quantities 
\begin{align}
	%
	%
	X & \equiv \cT + \Phi - \tilde{V} \cdot \tilde{N} \,, \\
	Y & \equiv \tb' \cdot \tn + \tn \cdot M' \cdot \tn \,, 
	\label{eq:Ydef}\\
	Z & \equiv \lp \tb' + \tn \cdot M' \rp \cdot \tb_\perp \,.
	\label{eq:Zdef}
\end{align}
That is, $ X$ and $ Y + Z$ determine respectively the Sachs-Wolfe and the integrated Sachs-Wolfe effect. By studying how $ X,Y$ and $ Z$ change under a gauge transformation we will automatically prove the gauge invariance of $ \Theta$. Indeed, in the next subsection we show that $ X$ transform as
\begin{align}
	X_{*}(x^{a}_{*}) & = X(x^{a}) + 
	\lc \alpha' + \xi' \cdot \lp \tn - \tb_{\parallel} - M \cdot n \rp \rc \Big|_{(x^{a})}\,,
\end{align}
with $ \tb_{\parallel} = \tb - \tb_{\perp} = \tn ( \tb \cdot \tn)$. Note that on the l.h.s the quantities are evaluated in the transformed path with coordinates $x^{a}_{*}$, while on the r.h.s quantities have to be evaluated along the (untransformed) path with coordinates $x^{a}$. In \refs{sec:GTisw} we show that 
\begin{align}
	\lp Y + Z \rp_{*} ( x^{a}_{*}) & = \lp Y + Z \rp ( x^{a}) + 
	\der{}{ \eta} 
	\lc \alpha' + \xi' \cdot \lp \tn - M \cdot \tn - \tb _\parallel \rp \rc\Big|_{(x^{a})} \,,
\end{align}
then a substitution into \eq{eq:ThetaXYZ} automatically proves the gauge invariance of the (logarithmic) temperature anisotropies, i.e., $ \Theta_{*} ( x^{a}_{*}) = \Theta ( x^{a})$.

\OLD{, that is, the temperature anisotropies are gauge invariant, as it has to be (since this is an observable). Let's point out some issues that will be important latter.}

\subsection{Gauge transformation of Sachs-Wolfe term $ X$}
\label{sec:GT-X}

The transformed $ X$ has the form 
\begin{align}
	X_{*}( x^{a}_{*}) = (\cT + \Delta \cT)+ (\Phi + \Delta \Phi) - ( \tilde{V} + \Delta \tilde{V}) \cdot (\tilde{N} + \Delta \tilde{N})\,,
\end{align}
where we wrote $ X_{*}( x^{a}_{*})$ to emphasize that $ X_{*}$ has to be evaluated in the transformed photon's coordinates $ x^{a}_{*}$. Now, using the definition of $ \un{ \tilde{N}}{i} = \un{\lp \tn + \tb_{\perp} + \tilde{V}_{\perp}/2 \rp}{i}$, we get 
\begin{align}
	\tilde{V} \cdot \Delta \tilde{N} + \tilde{N} \cdot \Delta \tilde{V} & = \un{\tilde{V}}{i} \lc \f12  \p{[i}\ \xi_{j]}\ \un{\tn}{j} + 
	\f12 (\xi'_{ \perp})_{i} \rc  + 
	\un{\tilde{N}}{i} \lc - \xi_{i}' + \xi^{c} \p{c} \unI{\tilde{V}}{i} + 
	\f12 \p{[i}\ \xi_{j]} \unI{\tilde{V}}{j} + M_{ij} \xi_{j}' \rc \notag \\
	& = \un{\tilde{N}}{i}\ \xi^{c} \p{c} \unI{\tilde{V}}{i} - \xi' \cdot \lp \tn + \tb_{\perp} - M \cdot \tn  \rp \,,
\end{align}
where in going from the first to the second line we used that at zero order $ \tilde{N} = \tn$. We remind the reader that we are not interested in quadratic terms in the gauge fields, so we do not consider the term $ \Delta \tilde{V} \cdot \Delta \tilde{N}$. Finally, $ \Delta \cT + \Delta \Phi  = \xi^{c} \p{c} \lp \cT  + \Phi \rp  + \alpha'- \tb \cdot \xi' $ so we get 
\begin{align}
	X_{*}(x^{a}_{*}) = \alpha' + \xi' \cdot & \lp \tn - \tb_{\parallel} - M \cdot \tn  \rp \notag \\
	& + \lp \cT + \Phi - \tilde{V} \cdot \tilde{N} \rp + 
	\lc \xi^{c} \p{c} \cT  + \xi^{c} \p{c} \Phi + (\xi^{c} \p{c} \unI{\tilde{V}}{i}) \un{\tilde{N}}{i}\rc 
	 \,,
	\label{eq:X*1}
\end{align}
with $ \tb_{\parallel} = \tb - \tb_\perp = \tn ( \tb \cdot \tn)$. We stress that the fields $ \alpha, \xi, \tb, M, \cT, \Phi $ and $ \tilde{V}$ must be evaluated at the coordinates  $ x^{a}_{*} = ( \eta, x^{i}_{*})$, with $ x^{i}_{*} = x^{i} - \lp \alpha\ \un{\tn}{i} + \xi^{i} \rp$. Then according to the discussion of \refs{sec:GT-nixi} we need to use \texttt{Rule 1}, that is, $ \xi^{c} \p{c} \to \lp \alpha\ \un{\tn}{i} + \xi^{i} \rp \p{i}$. In doing so, the second line of \eq{eq:X*1} is just
\begin{align*}
	\lp \cT + \Phi - \tilde{V} \cdot \tilde{N} \rp \Big|_{(x^{a})} = X(x^{a})\,,
\end{align*}
where we made a Taylor expansion. On the other hand, the first line of \eq{eq:X*1} is already linear in the gauge fields, so it is safe to evaluate theses terms along the path $ x^{a}$ (corrections will be quadratic in $ \xi^{a}$). Therefore we arrive at the final expression
\begin{align}
	%
	X_{*}(x^{a}_{*}) & = X(x^{a}) + \lc \alpha' + \xi' \cdot \lp \tn - \tb_{\parallel} - M \cdot \tn \rp \rc \Big|_{(x^{a})}\,.
	\label{eq:X*}
\end{align}
We want to emphasize the importance of the rule $ \xi^{c} \p{c} \to \lp \alpha\ \un{\tn}{i} + \xi^{i} \rp \p{i}$. Note that if we ignore it, then in the second line of \eq{eq:X*1} we would get a term of the form: $ \xi^{0} \p{0} \cT = \alpha \cT\,'$. We now argue on general grounds why this term cannot appear after a gauge transformation. Firstly, note that the gauge transformations of the metric perturbations Eqs.~\eqref{eq:GT-phi}-\eqref{eq:GT-M} are totally uncorrelated with $ \cT$ (the same happens for the gauge transformation of $ v$), so there is no possibility that an additional term inside $ \Theta$ cancels the term $ \alpha \cT\,'$. Secondly, if there is no time derivative of $ \cT$ before the gauge transformation (in the original $ \Theta$), how can it appear after the gauge transformation? Note that $ \cT$ is related to the temperature of the photon-baryon fluid, so $ \cT$ will depend on the microphysics describing the photon-baryon interaction. So unless we restrict to a
very particular case%
\footnote{One such particular case is to consider the limit of large scales of the CMB, and restrict to: adiabatic initial conditions and Einstein's field equations. In this specific case we know that $ \cT$ is totally determined by the metric perturbations, indeed we have $ \cT = - 2 \Phi/3 $.
} %
the metric perturbations (in general) will not contain enough information to compensate the term $ \cT\,'$.%
\footnote{We remind the reader that the Sachs-Wolfe formula \eq{eq:sw2nd-curved} is valid regardless of the nature field equations. For instance, it holds in modified theories of gravity. It is also valid regardless the nature of $ \cT$, that is, the photon-baryon temperature can contain any type of non-adiabatic perturbations.
} %
So again, if in the original $ \Theta$ there is no information on $ \cT\,'$, how is it that after the gauge transformation we can get such a term? The conclusion is: such a term is forbidden and to avoid it we need to apply the rule $ \xi^{c} \p{c} \to \lp \alpha\ \un{\tn}{i} + \xi^{i} \rp \p{i}$ in the transformation of $ X$. The not use of this rule led to a wrong result in \cite{Creminelli:2011sq} (see their Eq.27). In that paper the authors were interested in obtaining the squeezed limit of the CMB bispectrum. By arguing that a superhorizon perturbation (coming from adiabatic initial conditions during single-field-inflation) is locally equivalent to a coordinate transformation, they obtained a formula that is supposed to include corrections when the long-mode reenter the horizon. The error of this formula, introduced by the presence of the term $ \cT\,'$, was discussed and solved in \cite{Mirbabayi:2014hda}.

\OLD{See Appendix~\ref{app:} for a discussion on the relation between our approach and the one given in \cite{?}}

\subsection{Gauge transformation of $ Y$}

The gauge transformation of $ Y$ (see \eq{eq:Ydef}) is given by
%
%
\begin{align}
	Y( x^{a}_{*})  & = 
	\lp \tn + \Delta \tn \rp \cdot \lp \tb  + \Delta \tb \rp' + 
	\lp \tn + \Delta \tn \rp  \cdot \lp M + \Delta M \rp ' \cdot \lp \tn + \Delta \tn \rp \notag \\
	& = \lp \tn \cdot \tb' + \tn \cdot M' \cdot \tn \rp  + 
	J_{1} + J_{2} \,, 
	\label{eq:Y-J1J2}
\end{align}
where we have introduced $ J_{1} \equiv \tn \cdot \Delta \tb' + \tn \cdot \Delta M' \cdot \tn$ and $ J_{2} \equiv \lp \tb' + 2\, \tn \cdot M' \rp \cdot \Delta \tn$. Once more we discarded terms quadratic in the gauge fields, for instance $ \Delta \tn \cdot \Delta \tb$. In general, the fields $ \alpha, \xi, M, \tb$ appearing in the second line of the previous equation must be evaluated along the transformed path $ x^{a}_{*}$. In the case of $ J_{1}$ and $ J_{2}$ however, it is safe to evaluate them along the path $ x^{a}$ because these terms are already linear in the gauge fields (corrections will be quadratic in $ \xi^{a}$). So we have
\begin{align}
	J_{1}( x^{a}) & = \tn \cdot \lp - \p{}\alpha' + \xi'' \rp + 
	\lp \alpha'' - \un{\tn}{i}\, \un{\tn}{j}\, \p{i}\, \xi_{j}' \rp \notag \\
	& + \lb \xi^{c} \p{c} \lp \tn \cdot \tb + \tn \cdot M \cdot \tn \rp \rb' - 
	\lb ( \p{} \alpha + \xi') \cdot M \cdot \tn + ( \tn \cdot \tb)\, ( \tn \cdot \p{} \alpha) \rb' \notag \\
	 & + \lb \lc \f{( \tn \cdot \p{}) \xi - \p{} ( \tn \cdot \xi)}{2} - \xi' \rc \cdot \tb 
	 + \lc ( \tn \cdot \p{}) \xi - \p{} ( \tn \cdot \xi) \rc \cdot M \cdot \tn \rb' \,.
	\label{eq:J1}
\end{align}
Note that in each term involving curly braces, the direction vector $ \tn$ is a zero-order quantity, so the time derivative $ '$ only acts on the fields $ \alpha, \xi, M, \tb$. Now, from \eq{eq:deta} we know that along the photon trajectory we have $ \dd/ \dd \eta = \p{0} - \un{ \lp \tn + \tb + M \cdot \tn \rp }{i} \p{i}$\,. Therefore we can rewrite the first line of \eq{eq:J1} as %
\OLD{Now, we remind the reader that for the photon's trajectory we have $ q^{c} \equiv \dd x^{c}/ \dd \eta $. So $ \dd/ \dd \eta = q^{c} \p{c}  = \p{0} + q^{i} \p{i}$. Additionally, we know that up to first order $\un{\tn}{i} = - \lp \beta^i + q^{i} + M_{ij}\, \un{\tn}{j} \rp $, therefore we can rewrite the first line of \eq{eq:J1} as
}%
\begin{align}
	\tn \cdot \lp - \p{}\alpha' + \xi'' \rp + 
	\lp \alpha'' - \un{\tn}{i}\, \un{\tn}{j}\, \p{i}\, \xi_{j}' \rp = 
	\lp \dot{\alpha}' -  \tn \cdot \dot{\xi}' \rp  + \lp \tb + M \cdot \tn \rp \cdot \p{}  
	\lp \alpha' -  \tn \cdot \xi' \rp \,,
\end{align}
where a ``dot'' means total derivative w.r.t. $ \eta$. Additionally, using \eq{eq:GTni} we get
\begin{align}
	J_{2}  =  \lp \tb' + 2 \tn \cdot M' \rp \cdot \lp \p{ \perp} \alpha + 
	\f{ \p{} ( \tn \cdot \xi) - (\tn \cdot \p{}) \xi}{2}\rp  \,,
\end{align}
and putting everything together yields
\begin{align}
	J_{1} + J_{2} & = \lp \dot{\alpha}' + \tn \cdot \dot{\xi}' \rp  + 
	\lp \tb + M \cdot \tn \rp \cdot \p{}  
	\lp \alpha' + \tn \cdot \xi' \rp + 
	\lb \xi^{c} \p{c} \lp \tn \cdot \tb + \tn \cdot M \cdot \tn \rp \rb' \notag \\
	&  + \lc \p{ \perp} \alpha - \xi' - \tn \lp \tn \cdot \p{} \alpha \rp \rc \cdot \lp \tb' + M' \cdot \tn \rp  + J_{3}\,,
	\label{eq:J1+J2}
\end{align}
with
\begin{align}
	J_{3} & = \lc  - ( \p{} \alpha + \xi')'  + ( \tn \cdot \p{}) \xi' - \p{} ( \tn \cdot \xi') \rc \cdot M \cdot \tn  + \lc - \tn \lp \tn \cdot \p{} \alpha' \rp  + 
	\f{( \tn \cdot \p{}) \xi' - \p{} ( \tn \cdot \xi')}{2} - \xi''
	\rc \cdot \tb \notag \\
	& = - \lc \p{} \lp \alpha' + \tn \cdot \xi' \rp + \dot{\xi}' \rc 
	\cdot \lp M \cdot \tn + \tb \rp  + 
	\lc \p{ \perp} \alpha' + 
	\f{\p{} ( \tn \cdot \xi') - ( \tn \cdot \p{}) \xi'}{2} \rc \cdot \tb	\,.
	\label{eq:J3}
\end{align}
In the last line we used $ \dot{\xi}' = \xi'' - ( \tn \cdot \p{}) \xi'$, which is valid along the background trajectory. This is allowed because the term $ \xi' \cdot \lp M \cdot \tn + \tb \rp$ is already a second-order quantity so it can be evaluated along the unperturbed path (Born approximation). By replacing this expression for $ J_{3}$ we can simplify a bit \eq{eq:J1+J2}. However, let's first compute the gauge transformation of $ Z$ (see \eq{eq:Zdef}).

\subsection{Gauge transformation of $ Z$}

The transformed $ Z$ has the form 
\begin{align}
	Z_{*}( x^{a}_{*}) & = \lc \lp \tb + \Delta \tb \rp' + \lp \tn + 
	\Delta \tn \rp  \cdot \lp M + \Delta M \rp' \rc \cdot
	\lp \tb_\perp + \Delta \tb_\perp \rp \notag \\
	 & = \lp \tb' + \tn \cdot M' \rp \cdot \tb_\perp + J_{4}\,,
	\label{eq:Z-J4}
\end{align}
where $ J_{4} = \lp \tb' + \tn \cdot M' \rp \cdot \Delta \tb_\perp + \lc (\Delta \tb)' + \tn \cdot (\Delta \cdot M)' \rc \cdot \tb_\perp$. As always, we discarded terms which are quadratic in the gauge fields. The explicit form of $ J_{4}$ is 
\begin{align}
	J_{4} & = \lp \tb' + \tn \cdot M' \rp \cdot \lp - \p{_\perp}\alpha + \xi_\perp' \rp + 
	\lc \lp - \p{}\alpha' + \xi'' \rp - 
	\f{\p{} ( \tn \cdot \xi') + ( \tn \cdot \p{}) \xi'}{2} \rc \cdot \tb_\perp \notag \\
	& = \lp \tb' + \tn \cdot M' \rp \cdot \lp - \p{_\perp}\alpha + \xi_\perp' \rp  + 
	\dot{\xi}' \cdot \tb_{\perp} +  
	\lc - \p{ \perp}\alpha' + \f{( \tn \cdot \p{}) \xi' - \p{} ( \tn \cdot \xi')}{2} \rc \cdot \tb \,.
	\label{eq:J4}
\end{align}
We remind the reader that $ \p{}\alpha' \cdot \tb_{\perp} = \p{ \perp}\alpha' \cdot \tb_{\perp} = \p{ \perp}\alpha' \cdot \tb$. This is because, in the scalar product the notation $ \perp$ automatically kills the parallel part to $ \tn$. This is the reason why we changed $ \tb_{\perp}$ by $ \tb$ in the last term in \eq{eq:J4}. By using Eqs.~\eqref{eq:J1+J2}-\eqref{eq:J3} we easily arrive at
\begin{align}
	J_{1} + J_{2} + J_{4} & = \dot{\alpha}' + \dot{\xi}' \cdot \lp \tn - M \cdot \tn - \tb _\parallel \rp + 
	\lb \xi^{c} \p{c} \lp \tn \cdot \tb + \tn \cdot M \cdot \tn \rp \rb' \notag \\
	&  - \lc \xi'_{\parallel} + \tn \lp \tn \cdot \p{} \alpha \rp \rc \cdot \lp \tb' + M' \cdot \tn \rp \,,
	\label{eq:J1J2J4}
\end{align}
with $ \xi_{\parallel} = \xi - \xi_\perp = \tn ( \xi \cdot \tn)$. Additionally, note that we can write
\begin{align}
	\dot{\xi}' \cdot \lp \tn - M \cdot \tn - \tb _\parallel \rp & = 
	\der{}{ \eta} \lc \xi' \cdot \lp \tn - M \cdot \tn - \tb _\parallel \rp \rc - 
	\xi' \cdot \lp \dot{\tn}  - \dot{M}  \cdot \tn - \dot{\tb}_\parallel \rp \,.
	%
\end{align}
The time derivative of $ \tn$ can be obtained from \eq{eq:ni-1st-sol}, which yields
\begin{align}
	\dot{\tn} - \dot{M}  \cdot \tn - \dot{\tb}_\parallel = - 
	\tn \lp \tb' \cdot \tn + \tn \cdot M' \cdot \tn \rp + 
	\p{} \lp \tb \cdot \tn + \tn \cdot M \cdot \tn \rp\,.
	\label{eq:ndot}
\end{align}
With all the previous results, we can now obtain how the integrated Sachs-Wolfe effect transform under a gauge transformation.

\subsection{Gauge transformation of the ISW term $ Y + Z$}
\label{sec:GTisw}

The integrated Sachs-Wolfe is just the integral of $ Y + Z$, see Eqs.~\eqref{eq:Ydef}-\eqref{eq:Zdef}. Additionally, from Eqs.~\eqref{eq:Y-J1J2} and \eqref{eq:Z-J4} we get
\begin{align}
	\lp Y + Z \rp_{*} ( x^{a}_{*}) & = 
	\lp \tn \cdot \tb' + \tn \cdot M' \cdot \tn \rp \Big|_{( x^{a}_{*})} +
	\lp \tb' + \tn \cdot M' \rp \cdot \tb_\perp + 
	J_{1} + J_{2} + J_{4}\,.
	\label{eq:Y+Z}
\end{align}
The notation $ \Big|_{( x^{a}_{*})}$ is to remind the reader that both $ M$ and $ \tb$ have to be evaluated along the transformed path with coordinates $x^{a}_{*}$. From Eqs.~\eqref{eq:J1J2J4}-\eqref{eq:ndot} one obtains
\begin{align}
	J_{1} + J_{2} + J_{4} & = \der{}{ \eta} 
	\lc \alpha' + \xi' \cdot \lp \tn - M \cdot \tn - \tb _\parallel \rp \rc
	+ \lb \xi^{c} \p{c} \lp \tn \cdot \tb + \tn \cdot M \cdot \tn \rp \rb' \notag \\
	&  - \lp \tn \cdot \p{} \alpha \rp \cdot \lp \tn \cdot \tb' + \tn \cdot M' \cdot \tn \rp 
	- \xi' \cdot \p{} \lp \tb \cdot \tn + \tn \cdot M \cdot \tn \rp
	\,.
	\label{eq:J1J2J4-2}
\end{align}
Here, we encounter once more the subtle issue of gauge transformations on the CMB. It was discussed in \refs{sec:GT-nixi} that because of the relations $ \eta_{*} = \eta$ and $ x^{i}_{*} = x^{i} - \lp \alpha\, \un{\tn}{i} + \xi^{i} \rp $ a transformation rule is induced in the Sachs-Wolfe effect
\begin{align}
	\texttt{Rule 1:} \qquad \qquad 
	\xi^{c} \p{c} \to \lp \alpha\, \un{\tn}{i} + \xi^{i} \rp \p{i} \,,
\end{align}
that is, the time derivative $ \p{0}$ must be changed by a radial derivative, $ \p{r} = \un{\tn}{i}\, \p{i}$. That rule was also justified on physical grounds at the end of \refs{sec:GT-X}.

Calling $ f = \tn \cdot \tb + \tn \cdot M \cdot \tn$, we found a second rule which is needed in the integrated Sachs-Wolfe term:
\OLD{to prove the gauge invariance of $ \Theta$ is}
\begin{align}
	\texttt{Rule 2:} \qquad \qquad 
	\lp \xi^{c} \p{c} f \rp' \to \tn \cdot \p{} \lp \xi^{0} \p{0}\, f \rp + \lp \xi^{i} \p{i}\, f \rp'\,,
	\label{eq:rule2}
\end{align}
that is, once more we replace the time derivative by the radial derivative. Analogously to \texttt{Rule 1}, we can understand \eq{eq:rule2} on physical grounds. Indeed, note that if we ignore \texttt{Rule 2} then on the r.h.s of \eq{eq:rule2} we would get a term of the form $ (\xi^{0} \p{0} f)'$ and this quantity contains the term $ \alpha f''$. 

So we can ask: if there is no second time derivatives in the original ISW term, how they can appear after the gauge transformation? Note that unless we restrict to a very particular case, the second time derivatives of the fields $ M$ and $ \tb$ are in principle unrelated to $ M, \tb$ and its first-order derivatives. In order to relate $ M''$ and $ \tb''$ to $ M, \tb$ and its first-order derivatives, the field equations must be specified.%
\footnote{We remind the reader that the Sachs-Wolfe formula \eq{eq:sw2nd-curved} is valid regardless of the nature field equations. For instance, it holds in modified theories of gravity.
} %
Even if the field equations are specified, in general these equations will depend on additional variables as the energy-momentum tensor. So we can conclude that the metric perturbations (in general) will not contain enough information to compensate the term $ f''$.

Applying \texttt{Rule 2} to \eq{eq:J1J2J4-2} and replacing into \eq{eq:Y+Z} we arrive at
\begin{align}
	\lp Y + Z \rp ( x^{a}_{*}) & = 
	\lp \tn \cdot \tb' + \tn \cdot M' \cdot \tn \rp \Big|_{( x^{a}_{*})} +
	\lp \tb' + \tn \cdot M' \rp \cdot \tb_\perp + 
	\der{}{ \eta} 
	\lc \alpha' + \xi' \cdot \lp \tn - M \cdot \tn - \tb _\parallel \rp \rc\notag \\
	& + \alpha \lp \tn \cdot \p{} \rp \lp \tn \cdot \tb' + \tn \cdot M' \cdot \tn \rp 
	+ \xi \cdot \p{} \lp \tb' \cdot \tn + \tn \cdot M' \cdot \tn \rp 
	\,.
\end{align}
Finally, using that $ x^{i}_{*} = x^{i} - \lp \alpha\, \un{\tn}{i} + \xi^{i} \rp $, and by Taylor expanding the term $ \tn \cdot \tb' + \tn \cdot M' \cdot \tn$ we automatically obtain 
\begin{align}
	\lp Y + Z \rp_{*} ( x^{a}_{*}) & = \lp Y + Z \rp ( x^{a}) + 
	\der{}{ \eta} 
	\lc \alpha' + \xi' \cdot \lp \tn - M \cdot \tn - \tb _\parallel \rp \rc \,.
	\label{eq:YZ-final}
\end{align}
%
%

\subsection{Gauge invariance of the second-order Sachs-Wolfe formula $ \Theta$}

Remembering that the logarithmic temperature anisotropies are just
\begin{align}
	\Theta = X\Big|_{ \eta_o}^{\eta_e} + \int_{ \eta_e}^{ \eta_o} \dd\eta  \lp Y + Z \rp \,, 
\end{align}
and using Eqs.~\eqref{eq:X*} and \eqref{eq:YZ-final} we obtain $ \Theta_{*} ( x^{a}_{*}) = \Theta ( x^{a})$, that is, the temperature anisotropies are gauge invariant, as it has to be, since they are observables. With this result we finish our work.

\section{Conclusions}

We have introduced a cosmological frame which we called the river frame. This was motivated by realizing that the metric we introduced in \cite{Roldan:2017wvm} have many similarities with the so called Gullstrand-Painlev\'e metric used in the river model for black holes \cite{Hamilton:2004au,Hamilton:2017}. In the river frame, there is a background (the coordinate frame) with respect to which the river moves. Any other object can be thought as a fish swimming in that river. In particular, photons are fishes moving through the river with velocity $ c = 1$. By expressing quantities (e.g., the photon's four-momentum) in the river frame we have written the Sachs-Wolfe formula in a very intuitive and covariant form. Comparison of our model with the river model for black holes is given in \refs{sec:river-black holes}.

Then in \refs{sec:gt2nd}-\refs{sec:GInvariance} we have addressed the problem of gauge transformations on the CMB. We provided several compact formulas for the gauge transformations of the metric and fluid variables. Two rules (\texttt{Rule 1} and \texttt{Rule 2}) have been introduced which are necessary to deal with some subtle issues appearing when applying gauge transformations to the Sachs-Wolfe formula. All the time we expressed quantities in the river frame. Finally, we showed for the first time, the gauge invariance of the second-order temperature anisotropies as described by the second-order Sachs-Wolfe formula.

Proving the gauge invariance of the second-order Sachs-Wolfe formula and the introduction of the river frame, are the main results of this paper.

\section*{Acknowledgments}

I thank Miguel Quartin for useful discussions and suggestions. It is also a pleasure to thank to Cyril Pitrou for carefully reading the first version of the paper and making a lot of suggestions. Finally, I thank the anonymous referee for the very interesting comments and suggestions.

 \appendix

\section{Second-order gauge transformations}
\label{app:2ndGT}

In this appendix we compute the gauge transformations of the metric and fluid variables. The results obtained here are used in \refs{sec:gt2nd}-\refs{sec:GInvariance} to address the problem of gauge transformations as applied to the CMB and to prove the gauge invariance of the second-order Sachs-Wolfe formula.

\subsection{GT of the metric perturbations}

Writing the metric in two different ways will facilitate our calculations. 
\begin{align}
	\dd s^2  &  = a^2(\eta) e^{2\Phi} \lc - \dd \eta^2 + 
	2\, \beta_{j} \lp e^{-M} \rp^{j}_{\ i}\, \dd x^i \dd \eta 
	+ \lp  e^{-2M}\rp_{ij}\dd x^i \dd x^j \rc \,, \\
	& = a^2(\eta) \lc - e^{2\Phi} \dd \eta^2 + 2\, \omega_{i}\, \dd x^i \dd \eta 
	+ \lp  e^{-2N}\rp_{ij}\dd x^i \dd x^j \rc \,,
\end{align}
where in the second line we have introduced $ N = M - \Phi$, and $ \omega_i = \beta_{j} \lp e^{- M + 2 \Phi} \rp^{j}_{\ i}$\,. To obtain the transformation of the metric, we will use Eqs.~\eqref{eq:gt-rule1} and \eqref{eq:lieDer}, that is
\begin{align}
	g_{*} & = g + \f12 \lie \lp g + g_{*} \rp \,, \\
	\lie g_{ab} & = \xi^{c} \p{c}\, g_{ab} + \lp g_{ac}\, \p{b} + g_{cb}\, \p{a} \rp \xi^{c} \,.
\end{align}
%
%

\paragraph{Lapse perturbation}

By using $ (g_{*})_{00} = g_{00} + \f12 \xi^{c} \p{c} \lp g + g_{*} \rp_{00}  + \lp g + g_{*} \rp_{0c} \p{0} \xi^{c}$ we get
\begin{align*}
	2 \lp \Phi_{*} + \Phi_{*}^{2} \rp = 2 \lp \Phi + \Phi^{2} \rp & +  
	\lc \xi^{c} \p{c} \lp \Phi + \Phi_{*} \rp + 2 \cH \alpha \lp 1 + \Phi + \Phi_{*} \rp  \rc \\
	 & +  2 \lp 1 + \Phi + \Phi_{*} \rp \alpha' - \lp \beta + \beta_{*} \rp_{i} \xi_{i}' \,.
\end{align*}
From this we obtain at first order, $ \Phi_{*} = \Phi + \alpha'+\cH\,\alpha$, and then, substituting back into the previous equation yields 
\begin{align}
	\boxed{	\Phi_{*} = \lc \Phi + \alpha' +\cH \alpha \rc  + 
	\f12 \xi^{c} \p{c} \lp \Phi + \Phi_{*} \rp - \f12 \lp \beta + \beta_{*} \rp_i \xi_{i}'\,.}
	\label{eq:GT-phi-app}
\end{align}
We stress that (as we already said in \refs{sec:GT-metric}) in this work we will adopt the following notation: indices in $ \xi^{i}$ are raised and lowered with a delta Kronecker, that is $ \xi^{i} = \xi_{i}$. Although not necessary, this notation is convenient to get compact expressions.%
\footnote{As we will never need to use the covariant vector $ \xi_{a} = g_{ab} \xi^{b} $ no confusion should arise.
} %
%

\paragraph{Spatial part of the metric}

In the same way, using $ (g_{*})_{ij} = g_{ij} + \f12 \xi^{c} \p{c} \lp g + g_{*} \rp_{ij} + \f12 \lc \lp g + g_{*} \rp_{ic} \p{j} + \lp g + g_{*} \rp_{cj} \p{i} \rc \xi^{c}$ we get
\begin{align*}
	2 & \lp N_{*} - N_{*}^{2} \rp_{ij} = 2 \lp N - N^{2} \rp_{ij} + 
	\lc \xi^{c} \p{c} \lp N + N_{*} \rp_{ij} - 2 \cH \alpha \lp 1  - N - N_{*} \rp_{ij} \rc \\
	& - \lc \lp 1 - N - N_{*} \rp_{ik} \p{j} + \lp 1 - N - N_{*} \rp_{jk} \p{i} \rc \xi^{k}
	- \f12 \lc \lp \beta + \beta_{*} \rp_{i} \p{j} + \lp \beta + \beta_{*} \rp_{j} \p{i} \rc \alpha \,. 
\end{align*}
From this we obtain at first order, $ (N_{*})_{ij} = N_{ij} - \cH \alpha \delta_{ij} - \xi_{(i,j)}$, and then, substituting back into the previous equation yields 
\begin{align}
	& \lp N_{*}\rp_{ij} = \lc N_{ij} - \cH \alpha\, \delta_{ij} - \xi_{(i,j)} \rc  + 
	\f12 \xi^{c} \p{c} \lp N + N_{*} \rp_{ij} \notag\\
	& + \f14\lb \p{[i}\ \xi_{k]}\ \lp N + N_{*} \rp_{kj} - \lp N + N_{*} \rp_{ik} \p{[k}\ \xi_{j]} \rb
	 - \f14 \lc \lp \beta + \beta_{*} \rp_{i} \p{j} + \lp \beta + \beta_{*} \rp_{j} \p{i} \rc \alpha \,.
\end{align}
The notation $ (,)$ means symmetrization and $ [,]$ anti-symmetrization, that is, $ \xi_{(i,j)} \equiv (\p{i}\xi_{k} + \p{k}\xi_{i})/2$ and $ \p{[k}\ \xi_{j]} \equiv \p{i}\xi_{k} - \p{k}\xi_{i}$. Now, taking into account that $ M = N + \Phi$, we arrive at
\begin{empheq}[box=\widefbox]{align}
	& \lp M_{*}\rp_{ij} = \lc M_{ij} + \alpha'\ \delta_{ij} - \xi_{(i,j)} \rc  + 
	\f12 \xi^{c} \p{c} \lp M + M_{*} \rp_{ij} - \f12 \lp \beta + \beta_{*} \rp_{k} \xi_{k}'\ \delta_{ij} 
	\notag\\
	& + \f14\lb \p{[i}\ \xi_{k]}\ \lp M + M_{*} \rp_{kj} - \lp M + M_{*} \rp_{ik} \p{[k}\ \xi_{j]} \rb
	- \f12 \lp \beta + \beta_{*} \rp_{(i} \p{j)} \alpha \,.
\end{empheq}
%
%

\paragraph{Shift perturbation}

In the same way, from the gauge transformation of $ g_{0i}$ we get
\begin{align*}
	& \omega_{*}^{i} = \lc \omega_{i} - \alpha_{,i} + \xi_{i}' \rc +  
	\f12 \xi^{c} \p{c} \lp \omega + \omega_{*} \rp_{i} + \cH \alpha \lp \omega + \omega_{*} \rp_{i} \\
	& + \lc \f12\lp \omega + \omega_{*} \rp_{i} \p{0} - \lp \Phi + \Phi_{*} \rp \p{i} \rc \alpha + 
	\lc \f12\lp \omega + \omega_{*} \rp_{j} \p{i} - \lp N + N_{*} \rp_{ji} \p{0} \rc \xi^{j}\,.
\end{align*}
Now, using $ \beta_{i} = \omega^{j} \lp 1 + M - 2 \Phi \rp_{ji}$ which is valid up to second order, and taking the first-order expression $ \beta_{*}^{i} = \beta_{i} - \alpha_{,i} + \xi_{i}'$ which follows from the equation above, we obtain after some manipulations
\begin{align}
	\boxed{\beta_{*}^{i} = \lc \beta_{i} - \alpha_{,i} + \xi_{i}' \rc +  
	\f12 \xi^{c} \p{c} \lp \beta + \beta_{*} \rp_{i} + 
	\f12 \p{[i}\ \xi_{j]} \lp \beta + \beta_{*} \rp_{j} - 
	\f12 \lp \alpha_{,j} + \xi_{j}'\rp \lp M + M_{*} \rp_{ji}\,.}
\end{align}
%
%

\subsection{GT of the fluid perturbations}

\paragraph{Scalar field}

According to Eqs.~\eqref{eq:gt-rule1} and \eqref{eq:lieDer}, a scalar field $ S$ transform up to second order as
\begin{align}
		S_{*} = S + \f12 \xi^{c} \p{c} \lp S + S_{*} \rp \,.
\end{align}
Let's write $ S = \braket{S} e^{ \sigma}$, where $ \braket{S}( \eta)$ is the background value and $ \sigma$ gives the logarithmic anisotropies of $ S$. From the previous equation we obtain
\begin{align}
	\lp \sigma_{*} + \sigma_{*}^{2}/2 \rp = \lp \sigma + \sigma^{2}/2 \rp & +  
	\f12 \xi^{c} \p{c} \lp \sigma + \sigma_{*} \rp + 
	\f{ \alpha}{2} \lp 2 + \sigma + \sigma_{*} \rp  \p{0} \ln \braket{S}\,.
\end{align}
From this we get the first-order relation $ \sigma_{*} = \sigma + \alpha\, \p{0} \ln \braket{S}$. Substituting back in the previous equation yields 
\begin{align}
	\boxed{	\sigma_{*} =  \sigma + \alpha\, \p{0} \ln \braket{S} + 
	\f12 \xi^{c} \p{c} \lp \sigma + \sigma_{*} \rp \,.}	
\end{align}
Of particular interest is the density perturbation of cold dark matter which scale according to $ \braket{ \rho}_{m} \propto 1/a^{3}$ and so $ \p{0} \ln \braket{\rho}_{m} = - 3 \cH$.

\paragraph{Temperature of the baryon-photon}

For the baryon-photon fluid we write the temperature as $ T = \braket{T} e^{\cT}$, where $ \braket{T} \propto 1/a$, so we get
\begin{align}
	\boxed{\cT_{*} =  \cT - \cH \alpha + \f12 \xi^{c} \p{c} \lp \cT + \cT_{*} \rp \,.}
\end{align}
%
%

\paragraph{Four-velocity}

We are interested in obtaining the transformation rule of the \textit{peculiar velocity} (its tetrad components). The peculiar velocity is defined by the relation $ \com{u} = \gamma \lp u - v \rp$, which up to second order gives $ \un{v}{i} = \un{u}{i} - \beta^{i}$.
On the other hand, we know that (see \eq{eq:ui-u0}) 
\begin{align}
	\f{ \un{u}{i}}{ \un{u}{0}} = \f{ 1}{\beta^{0}} 
	\lp e^{-M} \rp^{i}_{\ j} \ \f{u^{i}}{ u^{0}}+ \un{ \tb}{i}\,.
\end{align}
Note that $ \un{u}{0} = \sqrt{1 + \un{u}{i}\ \unI{u}{i}}$ and $ \beta^0 \equiv \sqrt{1 + \beta_i \beta^i} $, so for second-order perturbations it is enough to take (in the previous equation) $ \un{u}{0} $ and $ \beta^{0}$ at zero-order. We also have that up to second order $ \un{ \tb}{i} = \beta^{i}$. Finally we get, up to second order
\begin{align}
	\un{v}{i} = \lp 1 - M \rp^{i}_{\ k}\ \f{u^{k}}{u^{0}} \,.
	\label{eq:v-2nd}
\end{align}
To obtain the gauge transformation of $ \un{v}{i}$, let's first compute the transformation of $ u^{k}/u^{0}$. From Eqs.~\eqref{eq:gt-rule1} and \eqref{eq:lieDer} we obtain for the four-velocity
\begin{align}
	u^{a}_{*} = u^{a} + \f12 \xi^{c} \p{c} \lp u^{a} + u^{a}_{*} \rp - 
	\f12  \lp u + u_{*} \rp^{c} \p{c}\ \xi^{a}\,,
\end{align}
from which follows that
\begin{align}
	\lp \f{u^{i}}{u^{0}} \rp _{*} = \f{u^{i}}{u^{0}} \f{ u^{0}}{u^{0}_{*}} + \f{1}{2 u^{0}_{*}} 
	\lc u^{0} \xi^{c} \p{c} \lp \f{u^{a} + u^{a}_{*}}{u^{0}} \rp + 
	\lp \f{u^{a} + u^{a}_{*}}{u^{0}} \rp \xi^{c} \p{c} u^{0} \rc - 
	\f12  \lp \f{u^{c} + u^{c}_{*}}{u^{0}_{*}} \rp \p{c}\ \xi^{a}\,.
\end{align}
\OLD{We write the four-velocity as $ u^{a} = u^{0} z^{a}$, so we are interested in the transformation rule of $ z^{i}$. From the previous equation we get}
Now, since $ u^{i}$ is at least first order, we just need $ u^{0}_{*} = u^{0} \lc 1 - \cH \alpha - \alpha' \rc $ which is valid up to first order. In the previous relation we used that $ u^{0} \propto 1/a $ at the background level. Then we obtain
\begin{align}
	\lp \f{u^{i}}{u^{0}} \rp _{*} = \f{u^{i}}{u^{0}} - \xi^{'i} + \f12 \xi^{c} \p{c} \lc \f{u^{i}}{u^{0}} + 
	\lp \f{u^{i}}{u^{0}} \rp _{*} \rc + 
	\f12 \lc \f{u^{j}}{u^{0}} + \lp \f{u^{j}}{u^{0}} \rp_{*} \rc \lp  \alpha \delta^{i}_{j} - \p{j}\, \xi^{i}\rp \,.
\end{align}
Substitution into \eq{eq:v-2nd}, and using the first-order relation $ \lp M_{*}\rp_{ij} = \lc M_{ij} + \alpha'\, \delta_{ij} - \xi_{(i,j)} \rc$  yields
\begin{align}
	\boxed{\un{v}{i}_{*} = \un{v}{i} - \xi^{'i} + 
	\f12 \xi^{c} \p{c} \lp \un{v}{i} + \un{v}{i}_{*} \rp + 
	\f12 \lp \un{v}{j} + \un{v}{j}_{*} \rp \f{ \p{[i}\ \xi_{j]} }{2} + 
	\f12 \lp  M + M_{*}\rp_{ij}  \xi^{'j} \,.}
\end{align}
The expressions given in this appendix will be used in \refs{sec:gt2nd}-\refs{sec:GInvariance} to discuss the gauge transformations as applied to the CMB.

\section{Why can we neglect the quadratic terms in $ \xi^{a}$?}
\label{app:quadratic_xi}

In this appendix we discuss in a more explicit way, why it was enough in \refs{sec:GInvariance} to consider second-order gauge transformations that are linear in the gauge parameter $ \xi^{a}$.

Firstly note that the second-order gauge transformation of an arbitrary geometrical quantity $ T$ can be thought as a quadratic function of the gauge field $ \xi^{a}$, in the sense that the transformed field $ T_{*}$ contains terms that are at most quadratic in $ \xi^{a}$. These quadratic terms are independent of the initial field $ T$ and can be obtained after two successive linear-in-$ \xi$ gauge transformations. Let's see how it works.

The fully second-order gauge transformation of the logarithmic intrinsic anisotropies is 
\begin{align}
	\cT_{*} (x_{*}) = \lc \cT - \cH \alpha + \xi^{c} \p{c} \lp \cT - \f12 \cH \alpha \rp \rc \Big|_{ x_{*} = x + \Delta x}  \,,
	\label{eq:cTquadratic-xi-1}
\end{align}
where $ x^{a}$ is independent of the gauge fields and $ \Delta x$ is linear in $ \xi$. Keeping only terms that are quadratic in $ \xi^{a}$ we have
\begin{align}
	\cT_{*} (x_{*}) \, \overset{q}{ = } - \lc (\Delta x)^{a} \p{a} \lp \cH \alpha \rp + \xi^{c} \p{c} \lp \f12 \cH \alpha \rp \rc \Big|_{ x }  \,,
	\label{eq:cTquadratic-xi-2}
\end{align}
where in going from \eq{eq:cTquadratic-xi-1} to \eq{eq:cTquadratic-xi-2} we Taylor expanded around $ x$. Note that we are using the notation $ \overset{q}{ = } $ to explicitly state that we are only considering the quadratic terms in $ \xi$.

Now consider two gauge fields $ \xi^{a}_{1}$ and $ \xi^{a}_{2}$, and let's apply two successive second-order gauge transformation keeping only terms that are linear in $ \xi^{a}_{1}$ and $ \xi^{a}_{2}$ respectively. So we have $ \cT \to \cT_{1*} \to \cT_{2*}$, where
\begin{align}
	\cT_{2*} (x_{2*}) & = \lc  \cT_{1*} - \cH \alpha_{2} + \xi^{c}_{2} \p{c} \cT_{1*} \rc \Big|_{ x_{2*} = x_{1*} + \Delta_{2} x} \notag \\
	& = \lc \lp \cT - \cH \alpha_{1} + \xi^{c}_{1} \p{c} \cT \rp - \cH \alpha_{2} + \xi^{c}_{2} \p{c} 
	\lp  \cT - \cH \alpha_{1} + \xi^{a}_{1} \p{a} \cT \rp \rc \Big|_{ x_{2*} = (x + \Delta_{1} x) + \Delta_{2} x}\,.
\end{align}
The term involving $ \xi^{c}_{2}, \xi^{a}_{1}, \cT$ can be neglected as it yields a third-order quantity. We remind the reader that the quadratic terms that we are looking for are independent of the fields $ \cT, \Phi$, etc., so it is enough to consider
\begin{align}
	\cT_{2*}(x_{2*}) & \overset{l}{ = } - \lc \cH \alpha_{1} + \cH \alpha_{2} + \xi^{c}_{2} \p{c} 
	\lp  \cH \alpha_{1} \rp \rc \Big|_{ x_{2*} = x + \Delta_{1} x + \Delta_{2} x} \notag \\
	& \overset{l}{ = } - \lc \Delta_{2} x \cdot \p{} \lp \cH \alpha_{1} \rp + \Delta_{1} x \cdot \p{} \lp \cH \alpha_{2} \rp + \xi^{c}_{2} \p{c} \lp  \cH \alpha_{1} \rp \rc \Big|_{x}\,,
	\label{eq:cT2*}
\end{align}
where we arrived at the second line by Taylor expanding around $ x$ and the notation $ \overset{l}{ = }$ means that we are only considering terms that are linear in $ \xi_{1}$ and $ \xi_{2}$. Now, note that by taking the limits: $ \xi_{1} \to \xi/\sqrt{2}$ and $ \xi_{2} \to \xi/\sqrt{2}$, the r.h.s of \eq{eq:cT2*} coincides with the r.h.s of \eq{eq:cTquadratic-xi-2}. That is, we have shown the terms that are quadratic $ \xi^{a}$ can be obtained by recursive linear-in-$ \xi$ GT. Although we only proved this for the specific case of $ \cT$, this can easily be shown for the other fields $ \Phi, \tilde{V}, M, etc$.

Finally, since the (logarithmic) temperature anisotropies $ \Theta$ is invariant under any linear-in-$ \xi$ gauge transformation, it will be invariant under recursive use of such linear-in-$ \xi$ GT. Consequently, $ \Theta$ is invariant under a full second-order GT.

\bibliographystyle{JHEP2015}
\bibliography{refs-up-to-1999,refs-2000-2009,refs-2010-2019}

\end{document}